\documentclass[11pt]{article}
\usepackage{epsfig,amscd,amssymb}
\newcommand{\be}{\begin{equation}}
\newcommand{\ee}{\end{equation}}
\newcommand{\bea}{\begin{eqnarray}}
\newcommand{\eea}{\end{eqnarray}}

\begin{document}

%
%
%
%
\title{Exact $S$-matrices for supersymmetric sigma models\\
and the Potts model}

\author{Paul Fendley$^1$ and Nicholas Read$^2$\\
\cr
$^1$ Department of Physics\\
University of Virginia\\
Charlottesville, VA 22904-4714\\
{\tt fendley@virginia.edu}
\\
\\
$^2$ Department of Physics,\\ Yale University,\\ P.O. Box
208120,\\
New Haven, CT 06520-8120\\{\tt nicholas.read@yale.edu}} \maketitle

\begin{abstract}
We study the algebraic formulation of exact factorizable
$S$-matrices for integrable two-dimensional field theories. We
show that different formulations of the $S$-matrices for the Potts
field theory are essentially equivalent, in the sense that they
can be expressed in the same way as elements of the Temperley-Lieb
algebra, in various representations. This enables us to construct
the $S$-matrices for certain nonlinear sigma models that are
invariant under the Lie ``supersymmetry'' algebras sl$(m+n|n)$
($m=1$, $2$, $n>0$), both for the bulk and for the boundary,
simply by using another representation of the same algebra. These
$S$-matrices represent the perturbation of the conformal theory at
$\theta=\pi$ by a small change in the topological angle $\theta$.
The $m=1$, $n=1$ theory has applications to the spin quantum Hall
transition in disordered fermion systems. We also find
$S$-matrices describing the flow from weak to strong coupling,
both for $\theta=0$ and $\theta=\pi$, in certain other
supersymmetric sigma models.
\end{abstract}


\section{Introduction}
\label{sec:intro}

In this paper we consider various models in statistical mechanics
and quantum field theory in two dimensions, some well-known and
some less familiar. These include the Potts model, restricted
solid-on-solid (RSOS) models, quantum spin chains, and nonlinear
sigma models (the latter two having Lie superalgebras as
symmetries). We focus here on the $S$-matrices of integrable
quantum field theories. The exact $S$-matrix for the massive
particles in an integrable quantum field theory can be found by
using a set of well-known criteria \cite{ZZ1}.  For example, the
multi-particle $S$-matrix elements factorize into sums of products
of two-particle elements. The consistency criterion for the
factorization is called the Yang-Baxter equation.

The main purpose of this paper is to show that such $S$-matrices can
be constructed, and the Yang-Baxter equation verified, by working with
elements of certain abstract algebras. Like any element of the
algebra, the $S$-matrix can be expressed in terms of the
generators. The $S$-matrices of various theories are images of the
same abstract $S$-matrix, in the representation appropriate to each
theory, under the homomorphism of the abstract algebra into the
algebra of linear maps on the vector space of internal states of the
particles in each theory. These $S$-matrices can conveniently be
written in terms of the generators of (the image of) the algebra in
each case. This allows us to use the ``same'' $S$-matrices to solve
different problems. Also, because one can often show that the
partition function of the model follows solely from the algebraic
properties of the $S$-matrix, there can be different representations
giving the same thermodynamics. This idea is very similar to that of
Temperley and Lieb, who studied lattice models algebraically within
the transfer matrix formulation, and introduced a different
representation of their algebra in order to find the partition
function \cite{TL}.

\begin{figure}
\begin{center}
\leavevmode \epsfxsize 0.25\columnwidth \epsffile{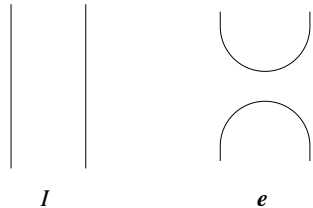}
\end{center}
\caption{The generators of the TL algebra} \label{fig:TL}
\end{figure}
\begin{figure}
\begin{center}
\leavevmode \epsfxsize 0.80\columnwidth \epsffile{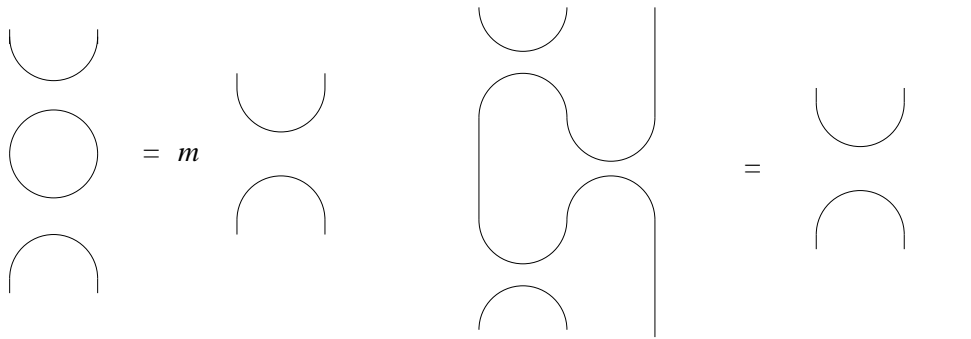}
\end{center}
\caption{The relations in the TL algebra} \label{fig:TLalg}
\end{figure}

The algebras that arise in the cases we study here are the
Temperley-Lieb (TL) \cite{TL} and Murakami-Birman-Wenzl (BMW)
\cite{BMW} algebras.  In fact, following old work in lattice
models \cite{Wadati,Martin}, we construct the SO$(3)$ version of the latter
within the former. As the TL algebra is of central importance, we
introduce it now. Its generators are labelled $e_i$, with $i=0$,
\ldots $2N-2$. They obey the relations \cite{TL,Baxter}
\bea e_i^2 &=& m e_i,\nonumber\\ e_i\, e_{i\pm 1}\, e_i &=&
e_i,\nonumber\\  e_i\,e_j&=&e_j\,e_i \quad (j\neq i,\;i\pm 1).
\label{tlrel}
\eea %
The relations (\ref{tlrel}) can be viewed as those arising from
non-intersecting lines: the identity $I$ and the generators $e_i$
acting on the two sites $i$, $i+1$ can be represented as in figure
\ref{fig:TL}, while the algebraic relations are shown in figure
\ref{fig:TLalg}. This non-intersection will be crucial in what
follows. (We have defined the TL algebra with an odd number of
generators, as this will be what we usually need, but the analog
with an even number of generators exists.)

The algebraic approach was already utilized in \cite{Smirpoly} to
study $S$-matrices proposed to describe O$(m)$-invariant field
theories for $|m|\le 2$ \cite{Zpoly}. The O$(m)$ $S$-matrices
were, strictly speaking, valid only for $m=1$ and $2$, although
they were cleverly written in a form which could seemingly be
applicable to continuous $m$. It was shown in \cite{Smirpoly} how
to formulate these $S$-matrices in terms of the TL algebra. Using
a different representation of this algebra gives the $S$-matrices
of the sine-Gordon model. This enabled the thermodynamics of the
O$(m)$ model in its high-temperature phase to be derived for any
$|m|<2$ \cite{FS}. This includes the interesting limit $m\to 0$,
where the model describes polymers (self-avoiding random walks).
In addition, yet another representation for the TL algebra yielded
the $S$-matrices for the $\phi_{13}$ perturbations of the minimal
models of conformal field theory \cite{ZRSOS}.

One of the key advantages of using the algebraic formulation of the
$S$-matrices (or that for the partition function of a lattice model)
is that it allows one to precisely define and calculate in theories
like this ``$m\to 0$'' limit. One of our main motivations for this
work was to be able to study the problem of percolation.  Percolation
is often defined as the $Q\to 1$ limit of the $Q$-state Potts
model. This limit can be defined precisely in the lattice model when
the TL (also known as the six-vertex, or XXZ chain) representation is
used (see section \ref{ssec:potts}), and in this representation the
continuum field theory at the critical point can be described by a
Coulomb gas \cite{dN,Nien,DF}. One can define the continuum field
theory off the critical point via perturbation theory, but the
algebraic formulation of the $S$-matrix is required to solve the
theory.

In order to describe the percolation limit $Q\to 1$, one needs an
$S$-matrix valid for all $Q$. Two seemingly different algebraic
formulations of the Potts $S$-matrix have been proposed in
\cite{Smir,CZ}.  The first was given in \cite{Smir}, where such an
$S$-matrix was found by studying the quantum affine algebra
$U_q(A_2^{(2)})$.  At special values of $Q$, this $S$-matrix
describes the integrable $\Phi_{21}$ perturbation of the minimal
models of conformal field theory \cite{BPZ}, described in section
\ref{ssec:min}. This formulation is precise, but this $S$-matrix
is quite intricate, and its intuitive connection to the original
Potts model is not very clear.  An elegant $S$-matrix for the
Potts model in the continuum limit was proposed in \cite{CZ}. Here
the connection to the original Potts representation of the lattice
model is quite clear: the world lines of the particles represent
domain walls between regions of different Potts spin states.
Unfortunately, this version of this $S$-matrix for arbitrary $Q$
is not precise: the algebra is not written out explicitly, and
the representations are defined only for $Q$ an integer. Moreover its
connection to the original $S$-matrix of \cite{Smir} is not clear.

In this paper, we show how the $S$-matrices of \cite{Smir,CZ} can
be written in terms of the same algebra, with elements in
different representations.  We then use this algebraic formulation
to find the $S$-matrix for another set of field theories. These
are nonlinear sigma models, where the fields take values on the
(super-) manifolds {\bf CP}$^{m-1}$ $\cong$
U($m$)/U$(1)\times$U$(m-1)$, or {\bf CP}$^{m+n-1|n}$ $\cong$
U($m+n|n$)/U$(1)\times$U$(m+n-1|n)$ in a version with sl$(m+n|n)$
Lie superalgebra symmetry (``supersymmetry''), as described in
section \ref{ssec:sigmod}. The actions of these models contain a
coupling constant $g_\sigma$ and $\theta$, the coefficient of the
topological (instanton) term. For $0<m\leq 2$, there is a critical
point at $\theta=\pi$ (mod $2\pi$), and the critical properties
are independent of $n$ for each $m$. If one perturbs the critical
theory by moving $\theta$ away from $\pi$, one obtains a massive
field theory. In this paper, we find the exact $S$-matrices for
these field theories when $m=1$, which corresponds to the
percolation case, and $m=2$, for general $n>0$.

A particular application of these results occurs in a disordered
noninteracting fermion model in two dimensions, in which there is
a spin quantum Hall transition \cite{kag,glr,smf,cardy,bcc},
similar to but distinct from the better-known integer quantum Hall
transition. The problem can be mapped to a nonlinear sigma model
on the target manifold OSp$(2n|2n)$/U$(n|n)$ \cite{az}, which is
$\cong$ ${\bf CP}^{1|1}$ for $n=1$. Indeed, a lattice version of
the problem \cite{kag} maps directly onto the corresponding
supersymmetric vertex model, and hence to percolation \cite{glr}.
Our results give the exact solution (that is, the $S$-matrices),
both for the bulk and the boundary, of the spin quantum Hall
transition in the scaling limit close to but {\em off}
criticality.

We begin in section \ref{sec:models} by reviewing the models to be
analyzed. In section \ref{sec:Potts}, we show that the two earlier
formulations of the $S$-matrix of the Potts model, though written
in different representations, are algebraically the same. We do
this by writing each $S$-matrix in terms of generators of the
$SO(3)$ BMW algebra. In section 4, we show that this algebra can
be represented within the TL algebra, thus giving another way of
describing the $S$-matrix. This makes it possible to write down
the $S$-matrices in the language of the sigma models. These
results also allow us in section 5 to characterize the boundary
$S$-matrix for free boundary conditions, originally given in
\cite{Chim}, and extend this also to the sigma models. This gives
an exact description of properties involving the edge states in
the spin quantum Hall effect. In section \ref{sec:disc}, we show
that other $S$-matrices that can be written in terms of the TL
algebra can also be applied to certain supersymmetric sigma
models, which hence are integrable on the lines $\theta=0$ and
$\theta=\pi$ as well as for the perturbation of $\theta$ away from
$\pi$, but perhaps do not apply to the ${\bf CP}^{m+n-1|n}$ sigma
models. In an appendix, we use our earlier results to study
$S$-matrices invariant under SU$(m)$ for $m>2$. In particular,
this allows us to find a unitary, crossing-symmetric $S$-matrix
for particles in the adjoint representation of SU$(m)$, which
however has some unphysical properties that prevent it from
representing the corresponding sigma models.


\section{The models and their phenomenology}
\label{sec:models}

In this section we introduce and review the various models to be discussed
in this paper.

\subsection{The Potts model}
\label{ssec:potts}

We begin with the $Q$-state Potts model, well-known in
two-dimensional classical statistical mechanics. It is defined by
placing a ``spin'' $s_i$ that takes values $1$, $2$,  \ldots, $Q$
($Q\geq 0$ a integer) at every site (or node) $i\in{\bf Z}^2$ of
the infinite square lattice (or graph), which we also call ${\bf
Z}^2$. The Hamiltonian is a sum of nearest neighbor terms only,
and is constructed to be invariant under the permutation (or
symmetric) group $S_Q$. This forces it to take the form %
$$ E_{\rm Potts} =
-\sum_{\langle ij\rangle} \delta_{s_i s_j}, $$ %
up to possible
multiplicative and additive constants. The $Q=2$ case is
equivalent to the Ising model. The Hamiltonian can be extended to
include ``magnetic field'' or source terms for the spin variables,
by adding the term $\sum_{i,s_i} h_{s_i}$, with $h_{s_i}$ a set of
parameters (the sum $\sum_{s_i}h_{s_i}$ is redundant for each $i$
and can be set to zero). Then correlation functions of arbitrary
functions of any set of spins $s_i$ can be calculated using
derivatives of $\ln Z$ with respect to the
$h_{s_i}$'s, where%
$$ Z=\sum_{\{s_i=1,\ldots,Q:i\in\Lambda\}} e^{-\beta E_{\rm
Potts}} $$%
and  the sum is over all configurations of the spins, and the
sites $i$ are restricted to a finite connected subset $\Lambda$ of
${\bf Z}^2$.

The partition function (with all $h_{s_i}=0$) can be rewritten
as %
$$ Z = \sum_{\{s_i=1,\ldots,Q:i=1,\ldots\}} \prod_{\langle
ij\rangle} \left( 1 + \delta_{s_i s_j} (e^\beta - 1)\right).$$%
The product can be multiplied out, and each term can be
represented as a (possibly disconnected) graph on the lattice by
drawing a line between sites $i$ and $j$ if $\delta_{s_i s_j}$ is
present in this term. For a given graph, the sum over spins can
now be easily done: it just results in a factor $Q^{n_c}$, where
$n_c$ is the number of connected
components (clusters) in the graph. The partition function is now %
\be Z=\sum_{\rm graphs} Q^{n_c} (e^\beta - 1)^{n_l},
\label{Zgraph}
\ee%
where $n_l$ is the number of links on the graph. In this form, $Q$
appears only as a parameter, and could be any complex number. We
take eq.\ (\ref{Zgraph}) to be the definition of the $Q$-state
Potts model for all $Q$. Strictly speaking, we have defined a
different statistical problem, in which the configurations are
graphs on the subset $\Lambda$ of the square lattice, and the
weight of each configuration is $Q^{n_c} (e^\beta - 1)^{n_l}$. The
partition function $Z$ is a starting point for obtaining
topological or geometric properties of the graphs. For example,
when $Q=1$ and $\beta\geq 0$, this describes bond percolation (the
probability of a bond being occupied being $1-e^{-\beta}$). When
$Q$ is a natural number, the graph formulation is equivalent to
the original Potts model, as far as thermodynamic (including the
simpler of the magnetic) properties---those that can be obtained
by differentiating $\ln Z$---are concerned. However, in general it
would be necessary to decorate the graphs if information about the
general Potts spin correlations as defined above were required.
These distinctions between formulations, though they may appear
unimportant, will be very relevant to our discussion, as we will
see. In the following, when we say that different models (or
different representations for the same transfer matrix or
$S$-matrices) are thermodynamically the same, we will mean the
extensive part of the scaling limit of $\ln Z$, in the absence of
the symmetry-breaking $h_{s_i}$ source terms, or their analogs in
other models.

\begin{figure}
\begin{center}
\leavevmode \epsfxsize 0.5\columnwidth
\epsffile{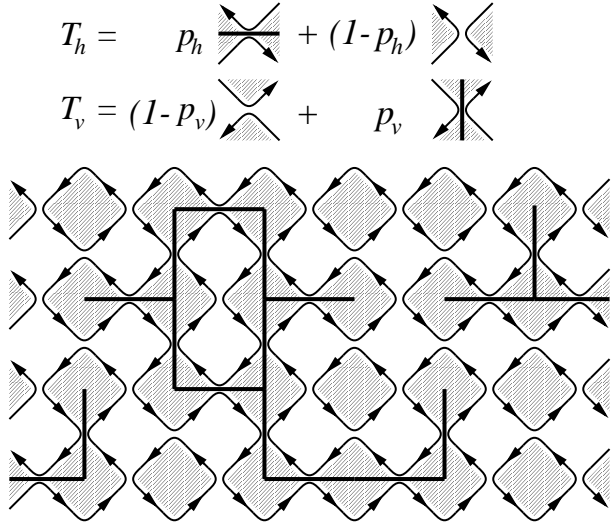}
\end{center}
\caption{Relation of transfer matrix, graphical expansion of the
Potts model, and loops on the medial graph} \label{fig:loops}
\end{figure}

The Potts model possesses a duality that exchanges large and small
$\beta$; the ferromagnetic (i.e.\ $\beta>0$) Potts model in the
thermodynamic limit, in which the square lattice becomes essentially
all of ${\bf Z}^2$ (with all $h_{s_i}=0$ from here on), has a phase
transition at its self-dual point $\beta=\beta_c$,
$e^{\beta_c}-1=\sqrt{Q}\geq0$, $Q\geq 0$. The model has long been
known to be solvable at this transition. As we will briefly review, it
can be mapped onto the solvable six-vertex model by using algebraic
techniques \cite{TL,Baxter} based on the transfer matrix approach, and
as a result, a number of interesting quantities can be computed
exactly \cite{Baxter}. One starting point for this mapping is the
graph formulation, eq.\ (\ref{Zgraph}). The set of graphs on the
square lattice is in one-to-one correspondence with a different
graphical problem, of configurations of non-intersecting loops that
fill the links of another square lattice (see Fig.\ \ref{fig:loops};
ignore the arrows on the loops for now). The loops are allowed to
touch at each node of the lattice, so there are two ways they can do
so at each node. This square lattice is the medial graph of the
original one; it has a node at the midpoint of each link of the
original lattice, joined by a link to another such node only if they
are on links of the original lattice that border a common face
\cite{Baxter}. The two possible configurations of lines at each node
of the medial graph determine the entire loop configuration. If the
loops at a node of the medial graph cut the link of the original
lattice, there is no line on that link in the original graph, while if
they do not, there is. This establishes the bijection between graphs
on the original square lattice, and non-intersecting loops filling the
medial square lattice \cite{Baxter}.

In the transfer matrix approach, the partition function is defined
(up to a numerical factor) using either a trace or an expectation,
depending on the desired boundary conditions, of a power of a
transfer matrix, which represents the addition of a row to the
lattice. The transfer matrix for the Potts model can be written
\be T\equiv T_1 T_3\cdots T_{2N-3}T_0T_2\cdots T_{2N-2}
\label{tltrans}\ee for the case of $N$ sites in a row of the Potts
model (or $2N$ sites of the model on the medial graph), and free
boundary conditions. The operators acting on medial-graph sites
$i$, $i+1$ are
\begin{eqnarray}
T_i&=&p_{\rm v} + (1-p_{\rm v}) e_i, \quad \hbox{($i$ even)}\nonumber\\
T_i&=&(1-p_{\rm h})+p_{\rm h}e_i, \quad \hbox{($i$ odd)},
\end{eqnarray}
in terms of the TL generators $e_i$ obeying (\ref{tlrel}) with
$Q=m^2$. (For the ferromagnetic Potts model one has $m>0$, but in
some formulations we can make sense of both signs of $m$.)  In the
original Potts model, the operators $e_{2i}$, $e_{2i+1}$ act on
the Potts spin variables $a_i=1$, \ldots, $Q$, for Potts sites
$i=0$, \ldots, $N-1$.  The $e_{2i}$'s act on the single site $i$,
while the $e_{2i+1}$'s act on the pair $i$, $i+1$ with both $i$,
$i+1$ in the range $0$, \ldots, $N-1$ (and on all other sites as
the identity in both cases). Explicitly, the matrix elements in
this representation, with $a_i'$ labelling rows, $a_i$ labelling
columns, are
\bea
(e_{2i})_{a_i',\,a_i}&=&1/m,\nonumber\\
(e_{2i+1})_{a_i'a_{i+1}',\,a_ia_{i+1}}&=&m\delta_{a_i\,a_{i+1}}
\delta_{a_i'\,a_i}\delta_{a_{i+1}'\,a_{i+1}}.\label{pottsrep}\eea
(The  $e_{2i}$ here is the $Q\times Q$ matrix with
all entries equal to $1/m$.) In the isotropic case,
$p_{\rm v}$ and $p_{\rm h}$ are equal, with
$$p_{\rm v}=p_{\rm h} ={e^\beta-1\over e^\beta-1+m}.$$
In the percolation case $Q=1$, $p_{\rm v}$ and $p_{\rm h}$ are the
probabilities for the occupation of bonds in, say, the vertical
and horizontal directions, respectively. In general, the
transition occurs at the self-dual point $p_{\rm v}=1-p_{\rm h}$,
which gives $e^\beta-1=m$ in the isotropic case.
In figure \ref{fig:loops}, we illustrate how the relation of the
graphical formulation and the nonintersecting loops formulations
look in terms of the TL generators.

Temperley and Lieb \cite{TL} realized that, because thermodynamic
properties (the partition function) can be calculated using only
the algebraic relations (\ref{tlrel}), any representation of the
TL algebra can be used. They found one in which each medial-graph
site $i$ carries a 2-dimensional vector space, whereas the Potts
representation above has a $Q$-dimensional space for each {\em
Potts} site. Thus the TL representation can be used for arbitrary
(even complex) $Q$; the resulting transfer matrix is that of the
six-vertex model.

In the anisotropic limit $p_{\rm h}\to 0$ with $(1-p_{\rm
v})/p_{\rm h}$ fixed, we can write $T\simeq e^{-p_{\rm
h}^{1/2}(1-p_{\rm v})^{1/2} H}$, where the Hamiltonian $H$ is
\be H=\epsilon\sum_{i\;{\rm even}} (1-e_i)
+\epsilon^{-1}\sum_{i\;{\rm odd}} (1-e_i),\label{tlham} \ee and
$\epsilon=\sqrt{(1-p_{\rm v} )/p_{\rm h}}$. Using the TL
representation, this is the Hamiltonian of the XXZ chain (here
with free boundary conditions) \cite{Baxter}. The Hamiltonian
chain and the transfer matrix $T$ can be solved by Bethe ansatz
methods, but only at the transition (self-dual) point, which in
the Hamiltonian chain becomes $\epsilon=1$ \cite{Baxter}; for
$\epsilon\neq 1$, one says that the nearest-neighbor coupling is
staggered. Similarly, in the six-vertex-model language, the
Boltzmann weights are staggered in space when $p_{\rm v}\neq
1-p_{\rm h}$.

One finds that this XXZ chain with $\epsilon=1$, or the six-vertex
model with $p_{\rm v}=1-p_{\rm h}$, is critical when $0<Q\le 4$,
and non-critical otherwise. Thus for $Q>4$, the transition through
$\epsilon=1$ (or $p_{\rm v}=1-p_{\rm h}$) in the Potts model is
first order, while for $0<Q\leq 4$, it is second order
\cite{Baxter}.


\subsection{The minimal models and the RSOS models}
\label{ssec:min}

For $Q\leq 4$, the Potts model near its critical point can be
described in the continuum by a massive field theory. Away from
the critical point, the Potts model maps onto a six-vertex model
where the Boltzmann weights are staggered from site to site. Such
a lattice model is not integrable, so none of the exact
computations in e.g.\ \cite{Baxter} are applicable here. However,
a lattice model not being integrable does not mean that the
corresponding field theory is not integrable. A famous example is
when a magnetic field is added to the Ising model at the critical
temperature. The field theory that describes the continuum limit
of this lattice model is integrable even though the lattice model
is not \cite{ZIsing}. The same is true for the Potts model off the
critical point. We will
describe this result in this section, but we need to first
introduce the restricted solid-on-solid (RSOS) models \cite{abf}.

The RSOS models are another set of well-known lattice models whose
transfer matrix at their critical points can be formulated in
terms of the TL algebra.  We will discuss in section
\ref{sec:Potts} how the RSOS representations of the TL algebra can
be used to construct $S$-matrices as well as lattice models.  In
this subsection we will discuss only the lattice models.  RSOS
models are defined by placing an integer variable called a height
on the sites of the lattice. The height variable is restricted to
take values from $1$, \ldots, $p$, and on adjacent sites it is
required to obey certain rules.  For example, in the original
Andrews-Baxter-Forrester (ABF) models
\cite{abf}, heights on adjacent sites must differ by $\pm 1$. The
conformal field theories describing the continuum limit of the
critical points arising in the ABF models are well-understood.
These theories are known as the minimal unitary models, and they
have central charge (the coefficient of the conformal anomaly)
$c=1-6/(p(p+1))$ \cite{BPZ}. More precisely, they are the A series
of modular-invariant conformal field theories, which also arise as
multicritical Ising models \cite{huse}. The case $p=2$ is trivial,
while the $p=3$ case is the Ising model, $p=4$ is the
tri\-critical Ising model, and $p=5$ case is the tetra\-critical
Ising model. The limit $p\to\infty$ limit gives the SU(2)$_1$ WZW
theory. These identifications require careful attention to the
role of periodic boundary conditions when the theories are
formulated on a torus, or on a study of the full set of local
operators in each theory.

Because their transfer matrices are identical when expressed in
terms of the TL generators, the critical point arising in an ABF
model is thermodynamically equivalent to that arising in the Potts
model when
\begin{equation}
Q=4\cos^2\left(\frac{\pi}{p+1}\right).
\label{Qp}
\end{equation}
When $Q$ given by this formula with $p$ integer is itself an
integer (i.e., when $p=3$, $5$, or $p\to\infty$), we can compare
the ABF models directly with the Potts model. For $Q=3$, $4$,
though the thermodynamics is the same, the critical theories are
not identical, but they are closely related (the three-state Potts
model yields one of the $D$ series of modular-invariant conformal
field theories, while the four-state Potts model is a $c=1$
orbifold theory).

For the TL (six-vertex, or XXZ) representation of the transfer
matrix, the corresponding critical field theory can be formulated
as a Coulomb gas \cite{dN,Nien}. In the language of conformal
field theory, this is a free boson with a charge at infinity
\cite{DF}. For generic $Q$, the conformal field theory is not
unitary, but when $Q$ takes the values in (\ref{Qp}) for $p$ an
integer $\geq 3$, it is possible to truncate or restrict (more
properly, project to a quotient space) the model and obtain a
nontrivial conformal field theory that is unitary. The restriction
can be done either in the conformal field theory, or in the
lattice model, and that is essentially what the ABF representation
of the TL algebra does---hence the name restricted solid-on-solid
model, as the six-vertex model is sometimes known as the
solid-on-solid model.

The continuum formulation involving conformal field theory is very
useful, because it allows us to define a massive field theory
describing the continuum limit of the Potts model near but not at
the critical point for any $Q$. Namely, one considers perturbing
the conformal field theory away from the critical point by adding
a relevant operator to the action. This is defined at least to all
orders in perturbation theory. The perturbing ``energy'' operator
corresponding to the Potts model is known as $\Phi_{21}$ in the
usual conventions of conformal theory \cite{BPZ,DF} (although in
\cite{DF} the subscripts are backwards from these now-accepted
conventions).  This definition is useful because it allows one to
show that this massive field theory is integrable \cite{ZIsing},
even though the underlying Potts lattice model is not.  This
integrable field theory has a factorized $S$-matrix that obeys the
YB equation, which we will discuss in detail in sec.\ \ref{sec:Potts}.
These considerations apply in any representation of the TL algebra
in the underlying lattice model.


\subsection{Spin chains and sigma models}
\label{ssec:sigmod}

Other models related to the TL algebra include SU($m$)-invariant
spin chains and vertex models. We take a chain of $2N$ sites
$i=0$, \ldots, $2N-1$, with an $m$-dimensional vector space at
each site. We consider the even sites to transform as the
fundamental (defining) representation of SU$(m$), and the odd as
the dual (here the same as the conjugate) of the fundamental; this
is the so-called $m$, $\overline{m}$ model. We then take the
nearest-neighbor interaction to be the unique (up to additive and
multiplicative constants) coupling that is invariant under this
action of SU($m$). It is essentially the invariant bilinear form
in the generators of SU($m$), one for each of the two sites, up to
similar constants again, and is the usual ``Heisenberg coupling''
of magnetism. Since the tensor product of $m$ and $\overline{m}$
representations decomposes into the singlet and the adjoint of
SU($m$), one can choose the constants such that the
nearest-neighbor interaction $e_i$ is $m$ times the projection
operator onto the singlet for sites $i$, $i+1$. Then the $e_i$'s
satisfy the TL algebra relations (\ref{tlrel}) \cite{BB,Affleck}.

The SU($m$) symmetry of the spin chains can be generalized to
the Lie superalgbra symmetry sl$(m+n|m)$ (supersymmetry for short).
The states of the supersymmetric chain
live in a graded vector space $V$ with
$m+n\geq 0$ even and $n\geq 0$ odd dimensions at each site with
$i$ even, and its dual $V^\ast$ at each site with $i$ odd (here
the dual is {\em not} the same as the conjugate representation)
\cite{RS}. We will give the details, as this is the basis for many
of our later results.
The space of states can be represented using boson and fermion
oscillators, with constraints. For $i$ even we have boson
operators $b_i^a$, $b_{ia}^\dagger$ such that
$[b_i^a,b_{jb}^\dagger]=\delta_{ij}\delta_b^a$ ($a$, $b=1$,
\ldots, $n+m$), and fermion operators $f_i^\alpha$,
$f_{i\alpha}^\dagger$ such that
$\{f_i^\alpha,f_{j\beta}^\dagger\}=\delta_{ij}\delta_\beta^\alpha$
($\alpha$, $\beta=1$, \ldots, $n$). For $i$ odd, we have similarly
boson operators $\overline{b}_{ia}$, $\overline{b}_i^{a\dagger}$ with
$[\overline{b}_{ia},\overline{b}_j^{b\dagger}]=\delta_{ij}\delta^b_a$
($a$, $b=1$, \ldots, $n+m$), and fermion operators
$\overline{f}_{i\alpha}$, $\overline{f}_i^{\alpha\dagger}$ with
$\{\overline{f}_{i\alpha},\overline{f}_j^{\beta\dagger}\}=
-\delta_{ij}\delta^\beta_\alpha$ ($\alpha$, $\beta=1$, \ldots,
$n$). Notice the minus sign in the last anticommutator; since our
convention is that the $\dagger$ stands for the adjoint, this
minus sign implies that the norms of any two states that are
mapped onto each other by the action of a single
$\overline{f}_{i\alpha}$ or $\overline{f}_i^{\alpha\dagger}$ have
opposite signs, and the ``Hilbert'' space has an indefinite inner
product (pedantically, it is the norm-squared that is negative).
The generators $e_i$ are Hermitian with respect
to the indefinite inner product. The special case $n=0$ is the
construction in Refs.\ \cite{BB,Affleck} with SU$(m)$ symmetry.

The supersymmetry generators are the bilinear forms
$b_{ia}^\dagger b_i^b$, $f_{i\alpha}^\dagger f_i^\beta$,
$b_{ia}^\dagger f_i^\beta$, $f_{i\alpha}^\dagger b_i^b$ for $i$
even, and correspondingly $-\overline{b}_i^{b\dagger}
\overline{b}_{ia}$, $\overline{f}_i^{\beta\dagger}
\overline{f}_{i\alpha}$, $\overline{f}_i^{\beta\dagger}
\overline{b}_{ia}$, $\overline{b}_i^{b\dagger}
\overline{f}_{i\alpha}$ for $i$ odd, which for each $i$ have the
same (anti-)commutators as those for $i$ even. Under the
transformations generated by these operators, $b_{ia}^\dagger$,
$f_{i\alpha}^\dagger$ ($i$ even) transform as the fundamental
(defining) representation $V$ of gl($n+m|n$),
$\overline{b}_i^{a\dagger}$, $\overline{f}_i^{\alpha\dagger}$ ($i$
odd) as the dual fundamental $V^\ast$ (which differs from the
conjugate of the fundamental, due to the negative norms). We
always work in the subspace of states that obey the constraints
\begin{eqnarray}
b_{ia}^\dagger b_i^a+f_{i\alpha}^\dagger f_i^\alpha &=& 1 \quad
(i\hbox{ even}),\label{constr1}\\ \overline{b}_i^{a\dagger}
\overline{b}_{ia}-\overline{f}_i^{\alpha\dagger}
\overline{f}_{i\alpha} &=& 1 \quad (i\hbox{ odd})\label{constr2}
\end{eqnarray}
(we use the summation convention for repeated indices of types $a$
or $\alpha$). These specify that there is just one ``particle''
(not a particle in the sense used elsewhere in this paper), either
a boson or a fermion, at each site, and so we have the tensor
product of alternating irreducible representations $V$, $V^\ast$
as desired. In the spaces $V^\ast$ on the odd sites, the odd
states (those with fermion number $-\overline{f}_i^{\alpha\dagger}
\overline{f}_{i\alpha}$ equal to one) have negative norm.

For any two sites $i$ (even), $j$ (odd), the
combinations
\begin{equation}
b_i^a \overline{b}_{ja} + f_i^\alpha\overline{f}_{j\alpha},\quad
\overline{b}_j^{a\dagger}b_{ia}^\dagger +
\overline{f}_j^{\alpha\dagger}f_{i\alpha}^\dagger \end{equation}
are invariant under gl($n+m|n$), thanks to our use of the dual
$V^\ast$ of $V$. The TL generators acting on sites $i$, $i+1$ with
$i$ even are then defined as %
\be e_i=(\overline{b}_{i+1}^{a\dagger}b_{ia}^\dagger +
\overline{f}_{i+1}^{\alpha\dagger}f_{i\alpha}^\dagger)(b_i^a
\overline{b}_{i+1,a} + f_i^\alpha\overline{f}_{i+1,\alpha}),
\label{TLsusyeven}
\ee
while for $i$ odd%
\be e_i=(\overline{b}_i^{a\dagger}b_{i+1,a}^\dagger +
\overline{f}_i^{\alpha\dagger}f_{i+1,\alpha}^\dagger)(b_{i+1}^a
\overline{b}_{ia} + f_{i+1}^\alpha\overline{f}_{i\alpha}).
\label{TLsusyodd}
\ee In
this case the model makes sense, and the $e_i$'s obey the TL
relations, for all integer $m$, though the $e_i$'s are $m$ times
the projector onto the singlet in $V\otimes V^\ast$ only for
$m\neq 0$ (the situation of interest in this paper). The TL
algebra relations depend only on $m$, not $n$, so we have
infinitely many distinct representations for each $m$. These
representations are naturally connected with the graphical
representation of the TL algebra shown in figs.\ 1 and 2, if we
view states in the representation $V$ as flowing along the lines
and remaining unchanged as they do so. Because of the
sl$m+n|n)$-invariant couplings, the lines can be consistently
oriented so that, say, $V$ flows in the direction along the arrow,
or $V^\ast$ in the reverse direction, as illustrated in fig.\ 3.

Using this any of these representations of the TL algebra, we may
again consider the Hamiltonian (\ref{tlham}), or alternatively the
transfer matrix (\ref{tltrans}). The partition function is then
that of the Potts model with $Q=m^2$. There is a transition at the
self-dual parameter values, which is first-order for $m>2$, and
second order for $m\leq 2$. The complete exact spectra of the
critical theories for $-2\leq m\leq 2$ have been determined
\cite{RS}. In particular, the case $m=1$ gives a sequence of
models that represent percolation, without an ill-defined limit
$Q\to 1$ being required.

There are nonlinear sigma models that are closely related to the
quantum spin chains/vertex models just discussed. These have target
manifolds {\bf CP}$^{m-1}$ $\cong$ U($m$)/U$(1)\times$U$(m-1)$, or
super-manifold {\bf CP}$^{m+n-1|n}$ $\cong$
U($m+n|n$)/U$(1)\times$U$(m+n-1|n)$.  The actions of these models
contain a coupling constant $g_\sigma$ and $\theta$, the coefficient
of the topological (instanton) term.  For $m>0$, the renormalization
group (RG) flow of $g_\sigma$ at weak coupling is towards strong
coupling. A phase transition occurs at $\theta=\pi$ (mod $2\pi$). At
$\theta=\pi$ the RG flow is towards a nontrivial critical theory
(conformal field theory) if $0<m\leq 2$. Different behavior occurs for
$m$ in the range $-2<m\leq 0$. In both the spin chains/vertex models
and sigma models, the exponents in the critical theories are
independent of $n$ for each $m$. For $m>2$ there are arguments that
the transition is first order, as originally found using the $1/m$
expansion \cite{oldCPN}. We refer to Ref.\ \cite{RS} and references
therein for the detailed arguments on all these theories and their
transitions.

The quantum spin chains discussed above can be generalized to
larger representations, labelled by their ``spin'' $\cal S$, with
${\cal S}=1/2$ in the original model. In the construction above,
this can be done by replacing 1 on the right hand side of Eqs.\
(\ref{constr1}), (\ref{constr2}) by the integer $2{\cal S}$, and
using the same Hamiltonian (\ref{tlham}) in terms of the $e_i$'s,
which are given by the same expressions, but no longer obey the TL
algebra. This Hamiltonian is again essentially the Heisenberg
nearest-neighbor coupling, now for all ${\cal S}$. As ${\cal S}\to
\infty$ a semiclassical mapping can be made onto the corresponding
nonlinear sigma models as just described \cite{Haldane}. In this mapping,
staggering the nearest neighbor coupling, as for $\epsilon\neq 1$,
changes the value of $\theta$ in the sigma model \cite{affleck}.
One then argues that the spin chains have transitions, when e.g.\
$2{\cal S}$ is odd and the nearest-neighbor couplings are
unstaggered, and for $m\leq 2$ these are second order and all in
the same universality class for each $m$, $n$. Hence the results
at ${\cal S}=1/2$ (the $V$, $V^\ast$ models described above) apply
to the critical points in all the spin chain models, and also in
the corresponding nonlinear sigma models as $\theta$ passes though
$\pi$ \cite{affleck,RS}. Thus the critical theories are closely
related to those in the Potts model at $Q=m^2$, $m$ integer. For
the case $m=2$, $n=0$, all this reduces to the well-known results
that the O$(3)$ sigma model at $\theta=\pi$ is critical, the
conformal theory is the SU$(2)_1$ WZW theory as in the spin-1/2
chain, and the $S$-matrices for the integrable perturbation by the
$p\to\infty$ limit of the $\Phi_{21}$ operator are that of the
sine-Gordon model at the coupling $\beta^2=2\pi$ \cite{ZZ1} (which
is equivalent to the $SU(2)_1$ WZW model perturbed by the operator
$\hbox{tr}\, g$.)


\section{The Potts $S$-matrix, algebraically}
\label{sec:Potts}

The minimal models perturbed by the $\phi_{21}$ operator were argued
to be integrable in \cite{ZIsing}. These field theories are
thermodynamically equivalent to the field theory describing the Potts
models near (but off) its critical point at the special values of $Q$
in (\ref{Qp}). The integrability was argued to persist for all values
of $Q$ in \cite{Smir}.

Once it is known a model is integrable, a variety of techniques can be
used to compute physical quantities. To do many such computations in
an integrable field theory, it is necessary to know the $S$-matrix.
In the $S$-matrix formulation, instead of dealing with a
two-dimensional classical finite-temperature model, or Euclidean
quantum field theory, we instead deal with a $1+1$-dimensional quantum
field theory obtained by continuing one of the coordinates of
two-dimensional spacetime to imaginary values.  It is almost always
simple to continue any physical quantity computed back and forth. The
$S$-matrix describes the scattering of massive particles in this
Lorentz-invariant quantum field theory. The theory is
Lorentz-invariant because the original lattice model is
rotationally-invariant in the continuum limit.

At integer values of $Q$, the Potts $S$-matrix is well understood. The
$S$-matrix of the Ising model ($Q=2$) for $\beta=\beta_c$ is trivial:
the only particle in the spectrum has $S=-1$. The particles for the
three-state model form a doublet under the $S_3$ symmetry; their
$S$-matrix is diagonal and was found in \cite{ZPotts}. The $S$-matrix
of the four-state model that of sine-Gordon model at $\beta^2=2\pi$,
whose $S$-matrix is given in \cite{ZZ1}. In this section we discuss
the Potts $S$-matrix at arbitrary values of $Q$.


\subsection{Chim-Zamolodchikov form}

A unified and physically-appealing way of understanding the Potts
$S$-matrices with $Q=2,3,4$ is given in \cite{CZ}.  In the
low-temperature phase ($\beta>\beta_c$), the Potts model has $Q$
degenerate vacua, and the $S_Q$ symmetry is spontaneously broken.
(We consider the dual high-temperature phase at the end of this
subsection.) One should be aware that at small non-zero
temperature, the thermodynamic pure phase consists mainly of Potts
spins in one of the $Q$ states, with typically small domains
within which one of the other $Q-1$ spin states occurs. Each of
these $Q$ symmetry breaking ensembles of spin configurations is
referred to as a vacuum in the quantum field theory point of view.
The elementary {\em excitations} of this quantum field theory
consist of kinks \cite{CZ}, at which location in space at a fixed
time the vacuum changes from one to another of the set of $Q$. The
worldlines of the kinks are thus domain walls in spacetime, but
note that in this sense, these domain walls are {\em not} present
in the vacua, if we assume a large system.

A typical excited state of the quantum field theory consists of
regions of the various vacua, separated by domain walls. The state at
a given time consists of a series of vacua $abcd\dots$,
where $a$, $b$, $c$, $d=1$, \dots, $Q$ with the requirement
that $a\ne b$, $b\ne c$, $c\ne d$, \dots. This configuration can then
be described as a set of kinks
$$|K_{ab}(u_1) K_{bc}(u_2) K_{cd}(u_3)\dots\rangle.$$
The kinks are assumed to be in sequence $1$, $2$, \ldots, from
left to right in position space. The variables $u_i$ are the
rapidities of the various kinks, defined in terms of the
energy and momentum of the kink
as $E=M\cosh u$ and $P=M\sinh u$, where $M$ is the
mass of the kink. One can say that there are then $Q(Q-1)$
different kinds of kinks, $K_{ab}$, with $a\ne b$ and $a$ the
vacuum to the left of the kink, $b$ the vacuum to the right.
However, it is more useful to say that there are
just $Q-1$ different kinds of particles, labelled by the possible
differences $a-b(\hbox{mod }Q)=1$, \dots, $Q-1$. For a given
vacuum $a$, one has $Q-1$ choices of what $b$ is. Hence the number
of $N$-kink configurations for $a$ at the left fixed is $(Q-1)^N$.
The $K_{ab}(u)$ can be thought of as (non-local) creation
operators applied to the vacuum, say $a$, the vacuum at infinity
at the left. We emphasize that the labelling of the kinks remains
sequential in scattering events, though the indices
can change, and the rapidities of two colliding particles are
interchanged (as usual).

\begin{figure}
\begin{center}
\leavevmode \epsfxsize 0.15\columnwidth \epsffile{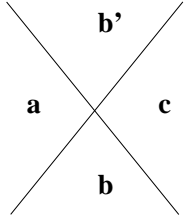}
\end{center}
\caption{Representing kink scattering by four vacua}
\label{fig:kink}
\end{figure}
Lorentz invariance requires that any two-particle $S$-matrix
elements be a function not of $u_1$ and $u_2$ separately, but only
of the combination $u\equiv u_1-u_2$. As a consequence of the
$S_Q$ symmetry, the $S$-matrix can be written in terms of four
matrices $A,B,C$ and $D$ \cite{CZ}
\begin{equation}
S_{\rm CZ}(u) = f^{(3)}(u) A + f^{(2)}(u) B + f^{(1)}(u) C +
f^{(0)}(u) D. \label{SCZ1}
\end{equation}
The matrix elements are labelled by four indices running from $1$
to $Q$, the four vacua involved in the scattering process, as in
figure \ref{fig:kink}.  The in state is $|abc\rangle$ and the out state is
$|a'b'c'\rangle$, but the matrix elements always vanish unless
$a=a'$, $c=c'$. The matrices $A$ and $C$ are diagonal, namely
\begin{equation}
 A_{a'b'c',\,abc}= \delta_{ac} \delta_{bb'} \delta_{aa'}\delta_{cc'},\qquad\qquad
 C_{a'b'c',\,abc} = (1-\delta_{ac})
 \delta_{bb'}\delta_{aa'}\delta_{cc'}.
\label{AC}
\end{equation}
The non-diagonal ones are
\begin{equation}
B_{a'b'c',\,abc} =
\delta_{ac}(1-\delta_{bb'})\delta_{aa'}\delta_{cc'}, \qquad\qquad
D_{a'b'c',\,abc}=(1-\delta_{ac})(1-\delta_{bb'})\delta_{aa'}\delta_{cc'}.
\label{BD}
\end{equation}
It is understood that all neighboring vacua must be different
(otherwise the particles would not be kinks). Thus any matrix
element with $a=b$, $b=c$, $a=b'$, or $c=b'$ vanishes. The
identity matrix $I$ is $A+C$.

CZ find the functions $f^{(i)}(u)$ by demanding that the
$S$-matrix satisfy the YB equation, and the constraints of
crossing and unitarity. Here we rewrite this result in a
different form. We define the matrices $E$ and $X$ by
\begin{eqnarray}
\nonumber
E&=& A+B,\\ X&=& A+B+C+D, \label{EX}
\end{eqnarray}
and write
\begin{equation}
S_{\rm CZ}(u)\propto g(u) I + h(u) E + X . \label{Scz}
\end{equation}
The proportionality $\propto$ here denotes equality up to a
numerical function of $u$; this factor will not be of interest in
this paper. Note that after allowing for this factor, the new
parameterization still contains one less independent function than
that in eq.\ (\ref{SCZ1}). The solution found by CZ does have this
form, and we hope to illuminate the reason for this in our
analysis. The functions $g(u)$ and $h(u)$ are found by requiring
that this $S$-matrix satisfy the YB equation. The YB equation
applies to three-particle states, labelled by $|abcd\rangle$. We
denote the two-particle $S$-matrix acting on the two-particle
state $|abc\rangle$ as $S_1$, and that acting on two-particle
state $|bcd\rangle$ as $S_2$. The YB equation is then the matrix
equation
$$S_i(u_1-u_2)S_{i+1}(u_1-u_3)
S_i(u_2-u_3) =S_{i+1}(u_2-u_3)S_i (u_1-u_3)S_{i+1}(u_1-u_2).$$

To solve the YB equation, we need to use various algebraic
relations satisfied by $E$ and $X$. Using their definitions, it is
simple to show that
\begin{eqnarray}
\nonumber
(E_i)^2 &=& (Q-1) E_i,\\
\nonumber
(X_i)^2 &=& (Q-2) X_i +  E_i.\\
X_i E_i &=& (Q-1) E_i.
\label{EXalg}
\end{eqnarray}
After a little work, one also finds that
\begin{eqnarray}
\nonumber E_iE_{i+1}E_i  &=& E_i,\\ \nonumber X_iE_{i+1}X_i &=&
X_{i+1}E_iX_{i+1},\\ \nonumber E_iX_{i+1}E_i  &=& (Q-1)E_i,\\ \nonumber
X_iE_{i+1}E_i  &=& X_{i+1}E_i,\\ \nonumber X_iX_{i+1}E_i  &=&
(Q-2)X_{i+1}E_i + E_i,\\ \nonumber X_iX_{i+1}X_i -  X_{i+1}X_iX_{i+1}
&=& X_{i+1}E_i + E_iX_{i+1} + X_i - E_i\\  &&
\mbox{}-X_iE_{i+1} - E_{i+1} X_i -  E_{i+1} +X_{i+1}.
\label{EXEalg}
\end{eqnarray}
All relations also hold with $i$ and $i+1$ interchanged, and with
the order of products in each term reversed. All generators here
and in the rest of the paper obey $A_i B_j=B_j A_i$ when
$|i-j|>1$. The relations involving only $E_i$ and $E_{i+1}$ are
those of the TL algebra, see eq.\ (\ref{tlrel}), but with $m$
replaced here by $Q-1$. Below we will see how the above algebra is
related to the SO$(3)$ BMW algebra.

By using the algebraic relations (\ref{EXalg},\ref{EXEalg}), the
YB equation reduces to a set of functional equations for $g$ and
$h$. One important thing to note is that to derive the functional
equations, one needs only to use the algebraic relations, and not
the explicit representations of $X$ and $E$. Since $Q$ appears
only as a parameter in the algebra, it appears only as a parameter
in these functional equations, which can then be solved for any
$Q$. These functional equations have been solved not only in
\cite{CZ}, but as we will explain below, in many other contexts as
well.  The solution is
\begin{eqnarray}
\nonumber g(u) &=& \frac{\sinh(\lambda u - 2i\gamma)}{\sinh(\lambda u)},
\\  h(u) &=& \frac{\sinh(\lambda u - i\gamma)} {\sinh(\lambda
u-3i\gamma)}, \label{gh}
\end{eqnarray}
where $\gamma$ is defined by
\begin{equation}
Q=4\sin^2\gamma \label{Qgamma},
\end{equation}
while $\lambda$ is unconstrained by the YB equation. The functions
CZ use are then easily found by rewriting $X$ and $E$ in terms of
$A,B,C$ and $D$; this yields
$$g+1 =\frac{f^{(1)}}{f^{(0)}},\qquad h+1 =\frac{f^{(2)}}{f^{(0)}},
\qquad g+h+1 =\frac{f^{(3)}}{f^{(0)}}.$$ The overall prefactor is
found by requiring that the $S$-matrix be unitary and
crossing-symmetric, as well as demanding consistency under the
bootstrap. These conditions are discussed in detail in
\cite{CZ,Dorey}. For Potts models, one finds  for example
\cite{CZ}
$$\lambda=\frac{3}{\pi}\gamma.$$

Another convenient way of writing this $S$-matrix is in terms of
``projectors'' ${\cal P}^{(t)}$, where $t=0$, $1$, $2$. We recall
that an idempotent in an algebra is an element, say $p$, that
obeys $p^2=p$, and we say two idempotents $p_1$, $p_2$ are
transversal if $p_1p_2=p_2p_1=0$. A set of transversal idempotents
that sum to the identity are projectors.  Note that these
definitions make no use of any inner product on the vector space,
and are entirely algebraic. Thus we have ${\cal P}^{(t)}{\cal
P}^{(t')}=\delta_{t,t'}{\cal P}^{(t)}$. Here $0$, $1$, and $2$
stand for the spin-0, -1, and -2 representations of $U_q({\rm
sl}_2)$, the connection with which will be explained below. We
define
\begin{eqnarray*}
&& {\cal P}^{(0)} = \frac{1}{Q-1}\big(A+B\big)=\frac{1}{Q-1}E,\\
&& {\cal P}^{(1)} = \frac{1}{Q-2}\big(C+D\big)=\frac{1}{Q-2}
\big(X-E\big),\\
&& {\cal P}^{(2)} = I-{\cal P}^{(0)} - {\cal P}^{(1)}.
\end{eqnarray*}
Substituting these operators into the CZ $S$-matrix, one finds
after some algebra that
$$ S _{\rm CZ}(u)\propto {\cal P}^{(2)} -
\frac{\sinh(\lambda u + 2i\gamma)}{\sinh(\lambda u - 2i\gamma)}
{\cal P}^{(1)} - \frac{\sinh(\lambda u + 3i\gamma)}{\sinh(\lambda
u - 3i\gamma)} {\cal P}^{(0)}.
$$
The unitarity of the $S$-matrices we discuss in this paper is
equivalent to $S(u)S(-u)=I$. The preceding relation makes this matrix
equation reduce to a functional relation for the prefactor.

Finally, we should mention the dual description. The above kink
description is very natural in the low-temperature phase. In the
high-temperature phase, the vacuum is $S_Q$ invariant. Because of
the symmetry, the fields or particles of the model should at least
include a set transforming in the standard $Q-1$-dimensional
representation of $S_Q$ (such a field appears in the
Landau-Ginzburg theory of the Potts model). This dimension is the
same as that for a kink in the low-temperature phase. This
observation provides the starting point for writing down the
$S$-matrix for these particles in this massive phase. In this
case, the matrices that take the place of $E$, $X$ for a pair of
particles must be $S_Q$ invariant. Accordingly, they should be
constructible within the TL algebra of such matrices in the
original Potts representation, acting on the $Q$ states for a
Potts spin. The exact formulas, which involve a projection that
cuts the space for these particles down to $Q-1$ dimensions, will
be given in section 3 below. An alternative point of view is that
we can describe the same low-temperature phase, but in terms of
the dual Potts spins, by interchanging the forms of the Potts
$e_i$'s for $i$ even and odd in eq.\ (\ref{pottsrep}).


\subsection{Smirnov form}

With all the above results on the CZ $S$-matrix, it is now
straightforward to prove that Smirnov's $S$-matrix \cite{Smir} can be
written in terms of the same algebra.  Precisely, we will show that
Smirnov's $S$-matrix also can be written in the form (\ref{Scz}),
where $X$ and $E$ obey the relations (\ref{EXalg},\ref{EXEalg}). As
discussed in the introduction, for the $S$-matrices as for the transfer
matrix of a lattice model, many physical quantities are the same when
the same algebra appears, albeit in different representations.  This
idea was described and demonstrated for the $S$-matrices of the
$\Phi_{13}$ perturbations of the minimal models in
\cite{Smirpoly}. There it was shown how the $S$-matrix for the $O(N)$
lattice model given in \cite{Zpoly} was in this sense equivalent to
the traditional RSOS $S$-matrix of \cite{ZRSOS}. Thus the content of
this subsection essentially consists of using the method of
\cite{Smirpoly} on the Potts/$\Phi_{21}$ $S$-matrices. This method
easily extends to the $\Phi_{12}$ $S$-matrices as well.

Smirnov finds a representation of the $S$-matrix for
all $Q$, and also shows that when [see eq.\ (\ref{Qgamma})]
$$\gamma = \pi \frac{p-1}{2(p+1)}$$ there is an RSOS kink
representation of $E$ and $X$ which results in a unitary
$S$-matrix. At these values of $Q$ (given by (\ref{Qp})) for $p$
an integer, the Potts model is thermodynamically equivalent to a
minimal model perturbed by the $\Phi_{21}$ operator. In
\cite{Smir}, it is argued that these theories can be viewed as
``restrictions'' of a field theory with symmetry under the quantum
affine algebra $U_q(A_2^{(2)})$. In our conventions, the parameter
$q$ of \cite{Smir}) is
$$q=-e^{2i\gamma},$$
so that
$$Q= (q^{1/2}+q^{-1/2})^2.$$
The starting field theory is non-unitary, but after
``restriction'' or ``truncation'', which amounts to taking the
quotient by null vectors in the non-unitary representations, the
quotient theory is argued to be unitary. It still possesses the
$U_q(A_2^{(2)})$ symmetry.

The $U_q(A_2^{(2)})$ symmetry algebra contains a $U_q(A_1)$ [i.e.,
$U_q({\rm sl}_2)$] subalgebra. In the unrestricted case, the space
of internal states for one particle is three-dimensional, and
corresponds to the the fundamental representation of $A_2$, or the
spin-1 representation of $U_q({\rm sl}_2)$. The particles in the
restricted theory are known as RSOS kinks. An RSOS kink transforms
as (a quotient of) the spin-1 representation of $U_q({\rm sl}_2)$,
or the corresponding $U_q(A_2^{(2)})$ representation. The spin-$s$
representations of $U_q({\rm sl}_2)$ are similar to those of
ordinary ${\rm sl}_2$, except that when $q$ is a root of unity,
representations with $2s\ge p+1$ are reducible. It is simplest to
first discuss RSOS kinks in the spin-$1/2$ representation, which
appear for example in the $\Phi_{13}$ perturbations of the minimal
models \cite{ZRSOS}. Spin-$1/2$ kinks interpolate between {\em
adjacent} wells of a potential with $p$ minima when $q^{p+1}=1$.
In other words, numbering the vacua $a=1$, $2$, \dots, $p$, the
kinks are of the form $|a\,b\rangle$, where $a$, $b$ are in $1$,
\ldots, $p$, and $b=a\pm 1$. (This is as distinguished from the
Potts kinks $|ab\rangle$ with $a\ne b$.) Spin-$1$ kinks have the
same allowed vacua, but with a different requirement. They are of
the form $|a\,b\rangle$ where $a$, $b$ are in $1$, \dots, $p$, and
$b=a\pm 2$, or $|a\, a\rangle$ for any $a=2$, \dots, $p-1$. One
can count the ``number'' $K$ of distinct particle states: while
the number of $N$-kink states is an integer for all $N$, as
$N\to\infty$, it behaves as $K^N$. Here it is straightforward to
show that (following e.g.\ \cite{pkondo})
$$K= 4 \cos^2\left(\frac{\pi}{p+1}\right) -1\leq 3.$$
Note that $K=Q-1$, just like the Potts kinks described above,
however, here we have a theory that makes sense for any $Q$ such
that $p$ is an integer, not just $Q=1$, $2$, $3$, $4$. For $Q=2$,
$3$, $4$, this construction reproduces the Potts model
construction of CZ.

The $S$-matrix (for both restricted and unrestricted theories) is
of the form
\begin{equation}
S_{\rm Sm}(u)\propto  (e^{-2\lambda u} - 1)e^{3i\gamma} B +
(e^{2\lambda u}-1)e^{-3i\gamma} B^{-1}+
4\sin(3\gamma)\sin(2\gamma), \label{Ssmir}
\end{equation}
where $B$ and $B^{-1}$ are $u$-independent matrices, and
$BB^{-1}=B^{-1}B=I$. In the unrestricted case, $B$ and $B^{-1}$
are $9$-dimensional. For both the unrestricted and restricted ($p$
integer) cases, the explicit forms can be found in \cite{Smir}. In
both cases, $B$ and $B^{-1}$ are associated with the constant
($u$-independent) solution of the YB equation for particles in the
spin-1 representation of $U_q({\rm sl}_2)$. This is why the
projectors discussed above are those of the spin-$0, 1$ and $2$
representations, the representations which appear in the tensor
product of two spin-1 representations of $U_q({\rm sl}_2)$.
However, $S_{\rm Sm}$ is not the standard solution of the
(rapidity-dependent) YB equation with spin-1 particles and
$U_q({\rm sl}_2)$ symmetry; it is instead associated with the
fundamental representation of $U_q(A_2^{(2)})$ \cite{IK,Jimbo}.
(The standard $U_q({\rm sl}_2)$-invariant $S$-matrix with spin-$1$
particles will be considered in section \ref{sec:disc}.)

\begin{figure}
\begin{center}
\leavevmode \epsfxsize 0.45\columnwidth \epsffile{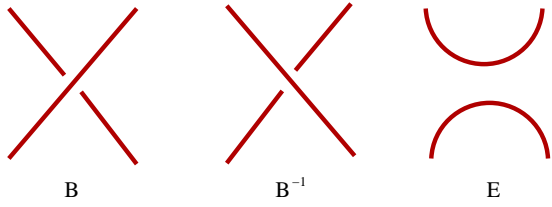}
\end{center}
\caption{The generators of the $SO(3)$ BMW algebra}
\label{fig:bmw}
\end{figure}
Now our task is to show that the $S$-matrix given by (\ref{Ssmir})
is equivalent to the $S$-matrix (\ref{Scz}), with the functions
$g$ and $h$ given by (\ref{gh}). We do this by relating $B$ and
$B^{-1}$ to $X$ and $E$. It was noted in \cite{Koubek} that $B$
and $B^{-1}$ satisfy the SO$(3)$ BMW algebra \cite{BMW}. This
algebra is usually written in terms of $B$ and a TL generator $E$,
where $B^{-1}$ is related to $E$ via $$B^{-1} = B +
(q-q^{-1})(E-I).$$ Pictorially, one can represent $B$, $B^{-1}$
and $E$ in figure \ref{fig:bmw}; $B$ is a braiding operation, while $E$ is
proportional to an idempotent (for $Q\neq 0$). The generators
$E_i$ then obey the TL algebra (the first relation in each of equations
(\ref{EXalg}) and (\ref{EXEalg})), where indeed
$Q=(q^{1/2}+q^{-1/2})^2$. The remaining relations of the SO$(3)$
BMW algebra  are
\begin{eqnarray}
 \nonumber B_iE_i &=& q^{-2} E_i,\\ \nonumber B_iB_{i+1}B_i  &=&
B_{i+1}B_iB_{i+1},\\ \nonumber B_iE_{i+1}B_i  &=&
B^{-1}_{i+1}E_iB^{-1}_{i+1},\\ \nonumber B_iB_{i+1}E_i  &=&
E_{i+1}B_{i}B_{i+1}  = E_{i+1}E_i,\\ \nonumber B_iE_{i+1}E_i  &=&
B^{-1}_{i+1}E_i ,\\  E_iB_{i+1}E_i  &=& q^2 E_{i}
,\label{EBalg}
\end{eqnarray}
and those relations given by interchanging $i$ and $i+1$, as well
as reversing the order of both sides.
It is now simple to show
that these relations are identical to (\ref{EXalg}) and
(\ref{EXEalg}), if
\begin{eqnarray}
\nonumber
B&=& qI - X + q^{-1}E,\\ B^{-1}&=& q^{-1} I - X + qE.
\end{eqnarray}
Plugging these into (\ref{Ssmir}), one indeed recovers (\ref{Scz})
and (\ref{gh}).

We now have a unified form of the $S$-matrix for the scaling limit
of the $Q$-state Potts model in the region $Q\le 4$. It is given
by eqs.\ (\ref{Scz}) or (\ref{Ssmir}), in terms of generators of
the SO$(3)$ BMW algebra and thus, as an element in the algebra,
independent of any representation. The YB equation has been
verified, and the functions in eqs.\  (\ref{Scz}) or (\ref{Ssmir})
found, using only the relations in the algebra. Consequently, by
using a representation of the BMW algebra, in which the generators
become matrices, we obtain the $S$-matrix appropriate for that
representation (up to overall functions independent of the
representation). There are four separate representations for this
$S$-matrix, with two more to be described in the next section. The
first two we discussed are the low-temperature Potts
representation \cite{CZ}, and its dual, the high-temperature Potts
representation; both are valid for $Q=2$, $3$, and $4$. The third
is the unrestricted $U_q(A_2^{(2)})$-invariant $S$-matrix, valid
for all $Q\leq 4$. Finally, the fourth is the restricted
$U_q(A_2^{(2)})$ or RSOS kink representation, valid for $p$
integer $\geq 3$, where $Q=4\sin^2\gamma$ and
$\gamma=\pi(p-1)/[2(p+1)]$ \cite{Smir}. The last two versions are
self-dual, in the sense that the matrices for $e_i$ are the same
for $i$ even and odd.

We have again omitted the overall function multiplying the
$S$-matrix; this is determined by crossing, unitarity and the
bootstrap; it can be found in either \cite{Smir} or \cite{CZ}
(some subtleties regarding the bound states are discussed in
\cite{Dorey}). Using the fact that $E$ and $X$ are Hermitian (with
respect to the appropriate inner product) in all the
representations we discuss, all these conditions can be reduced to
functional equations by using only the algebraic relations. For
$Q\le 3$, the particles discussed so far (corresponding to spin-1
kinks in $U_q({\rm sl}_2)$) are the only ones, while for $3<Q\leq
4$, the spectrum contains also a stable singlet particle (i.e.\
corresponding to spin 0 in $U_q({\rm sl}_2)$). There should be no
difficulty in writing down the spin 0-0, 0-1, and 1-0
$S$-matrices, and verifying the YB equation, in terms of the BMW
algebra also. By redefining the parameters $q$ and $\gamma$ as
described in \cite{Smir}, all of the above considerations also
apply to the $\Phi_{12}$ perturbations of the minimal models of
conformal field theory. In the language of lattice statistical
mechanics, these correspond to the scaling limit of the
tricritical $Q$-state Potts models.


\section{The Potts $S$-matrix in TL language and the sigma models}

In this section we provide yet another way of representing the
Potts $S$-matrix. This representation is entirely in terms of TL
generators. There are several reasons why this is useful. It is
somewhat simpler to deal with the TL algebra only, instead of the
much more complicated BMW algebra. This provides a nice intuitive
connection with the Boltzmann weights of the lattice model, which
also can be expressed in terms of generators of the TL algebra.
This representation also illuminates a connection to the
SU$(m)$-invariant representation with $Q=m^2$, described in the
introduction; in fact, we must generalize to sl$(m+n|n)$
supersymmetry in order that this be meaningful for $m=1$.
According to the arguments of Refs.\ \cite{glr,RS}, this describes
the supersymmetric sigma models on ${\bf CP}^{m+n-1|n}$, $m=1$,
$2$, with the critical theory at $\theta=\pi$  perturbed by
changing $\theta$. Finally, it will also allow us to provide in
the next section a nice interpretation of the boundary
$S$-matrices of \cite{Chim}, with applications to all the above
representations.


\subsection{The Potts $S$-matrix in terms of TL generators}

\begin{figure}
\begin{center}
\leavevmode \epsfxsize 0.30\columnwidth \epsffile{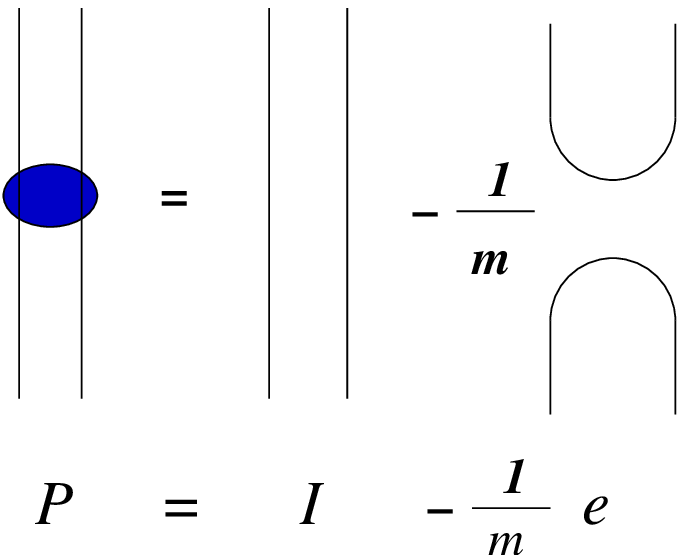}
\end{center}
\caption{The projection onto the adjoint}
\label{fig:proj}
\end{figure}

Rather than immediately giving the formulas for the SO$(3)$ BMW
generators in terms of the TL generators, we will begin by briefly
indicating the motivation that led to them. First, we note that in
the Smirnov form of the $S$-matrix, the possible states of a
single particle (kink) were related to the spin-$1$ representation
of $U_q({\rm sl}_2)$. Spin-$1$ can be obtained within the tensor
product of two spin-$1/2$ $U_q({\rm sl}_2)$ representations. In
iterated tensor products of spin-1/2, the TL algebra is the
commutant of $U_q({\rm sl}_2)$, the algebra of invariant operators
in the representation, at least for generic $q$ (see Ref.\
\cite{cp} for a review). Thus the projectors of two-particle
states $i$, $i+1$ onto spin-0 and spin-1 are given by $e_i/m$,
$I-e_i/m$, respectively. This is displayed pictorially in fig.\
\ref{fig:proj}. We then expect that the $U_q({\rm
sl}_2)$-invariant $S$-matrix of spin-1 particles can also be
written in terms of TL generators.

\begin{figure}
\begin{center}
\bigskip\bigskip
\leavevmode \epsfxsize 0.55\columnwidth \epsffile{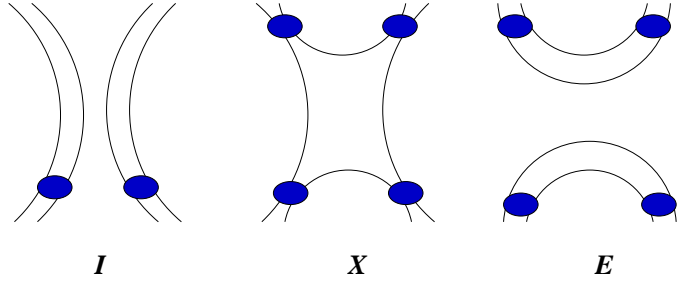}
\end{center}
\caption{The generators of the BMW algebra in terms of the TL
algebra} \label{fig:bmwtl}
\end{figure}

To find the explicit form, it is useful to consider the graphical
representation of the TL algebra, already illustrated in figs.\ 1
and 2.
Essentially, the TL algebra is the algebra of diagrams consisting
of (topological equivalence classes of) non-crossing lines that
join $4N$ points on a circle in pairs through the interior, the
product being joining of the $2N$ points on a semicircle at the
bottom of one diagram to the $2N$ on a semicircle at the top of
another. Each point represents a spin-1/2 of $U_q({\rm sl}_2)$.
The generators $e_i$ represent a pair of $180^\circ$ turns.
Including the projector for each pair of spin-1/2's to spin-1,
then the scattering of kinks represented by a pair of lines should
be described by the possible ways of either continuing the lines
through the scattering, or joining them in $180^\circ$ degree
turns, but never allowing the lines to cross or join their
partner. There are exactly three linearly-independent ways to do
this, just as there are three terms in the decomposition of the
tensor product of spin-1 with itself. We arrive at the
representation of the three possibilities shown in figure
\ref{fig:bmwtl}. These are (up to constant factors) $I$ (the
identity), $E_i$, and $X_i$ defined earlier (and this was the
reason for those definitions). The projectors are linear
combinations of $I$ and $e_i$'s, but do not introduce crossings
either. This construction has been known for some time
\cite{Wadati,Martin,saleur}.

The preceding pictures translate into the following expressions.
In terms of the projectors ${\cal P}_i=I-e_i/m$ onto spin-1, we
have
\begin{eqnarray}
\nonumber
E_{i} &=& {\cal P}_{2i+1}\, {\cal P}_{2i+3}\, e_{2i+2}\,
e_{2i+1}\, e_{2i+3}\, e_{2i+2}\, {\cal P}_{2i+1}\, {\cal
P}_{2i+3},\\  X_{i} &=& m{\cal P}_{2i+1}\, {\cal P}_{2i+3}\,
e_{2i+2}\, {\cal P}_{2i+1}\, {\cal P}_{2i+3}. \label{EXTL}
\end{eqnarray}
Using only the TL algebra relations (\ref{tlrel}), one can show
(after some effort) that the generators $E_i$ and $X_i$ defined by
(\ref{EXTL}) satisfy the relations (\ref{EXalg}), (\ref{EXEalg})
\cite{saleur}. In this proof, a very useful set of identities is
$$E_{i}  = {\cal P}_{2i+2\pm1}\, e_{2i+2}\, e_{2i+1}\, e_{2i+3}\,
e_{2i+2}\,{\cal P}_{2i+2\pm1}\,  = {\cal P}_{2i+2\pm1}\,
e_{2i+2}\, e_{2i+1}\, e_{2i+3}\, e_{2i+2}\, {\cal P}_{2i+2\mp1}\,
.$$ This identity is obvious in the graphical language, but takes
some work to prove using the TL algebra alone.

Thus from any representation of the TL algebra we can find using
eq.\ (\ref{EXTL}) a representation of the SO$(3)$ BMW algebra, and
hence of the Potts $S$-matrix. The original lattice
Potts model representation (\ref{pottsrep}) of the TL algebra
yields the low-temperature Potts kink $S$-matrix of CZ, while
the same Potts representation but with $-1$ added to the indices
on the right-hand side in eq.\ (\ref{EXTL}) reproduces the dual
high-temperature Potts $S$-matrix mentioned at the end of
subsection 2.1. In these cases, the projectors ${\cal P}_{2i}$ or
${\cal P}_{2i+1}$ are onto a space of $Q-1$ states for each
particle. The six-vertex/XXZ representation of the TL algebra
should result in the unrestricted $U_q(A_2^{(2)})$-invariant
$S$-matrix, while the RSOS spin-1/2 kink representation of
the TL algebra should result in the RSOS spin-1 kink $S$-matrix
given in \cite{Smir}, as discussed in subsection 2.2.


\subsection{The $S$-matrices for the supersymmetric sigma models}
\label{ssec:ssS}

The final examples we want to discuss in a little more detail are
the supersymmetric sigma models with sl$(m+n|n)$ supersymmetry.
We find here the $S$-matrix describing the massive field theory
given by perturbing the topological angle $\theta$ away from
$\pi$. All values of
$\theta$ will be mod $2\pi$ here; later we discuss
the more general situation.
This $S$-matrix is valid for
$m=1$, $2$, $n>0$ (the cases $n=0$ are trivial and known,
respectively), and the related lattice models.
In the appendix we will discuss the situation with
$m>2$.

A useful starting point is the
result in the $1/m$ expansion of the ${\bf CP}^{m-1}$ sigma models
\cite{oldCPN} (also valid for the sl$(m+n|n)$-invariant
extensions) that the transition for bringing $\theta$ through $\pi$
is first order. The first-order transition point is
believed to persist for all $m>2$. At the transition point, there are
two competing phases (vacua) of the same energy density (energy
per unit length), that correspond to, say $\theta=\pi-0^+$ and
$\pi+0^+$, respectively, where $0^+$ represents a positive
infinitesimal. Elementary excitations are domain walls (kinks)
separating these phases, which are simply the boson excitations or
``spinons'' of the theory, that transform in the fundamental $V$
of SU($m$), and the dual $V^\ast$ for the antiparticles. For the
kinks in the fundamental, we have the $\theta=\pi-0^+$ vacuum to
the left of it, and the $\theta=\pi+0^+$ vacuum to the right; the
reverse applies to the dual fundamental. Thus if kinks are
present, they must alternate in type along the system. Otherwise
there will be a region of yet another phase in between, with a
larger energy density. Off the transition point $\theta=\pi$, the
energy density for the two phases is different. Then the kinks are
confined in pairs by a linear potential, and transform in the
representation $V\otimes V^\ast$ of SU($m$) or sl$(m+n|n)$. In the
phase $\theta$ slightly $<\pi$, the pair has $V$ to the left of
$V^\ast$, and the reverse in the phase $\theta$ slightly $>\pi$.
For $0<m\leq 2$, the transition is second order, but the picture
of the phases off the transition remains valid. The field theories
we construct via their $S$-matrices describe either of these two
phases, in the scaling limit near $\theta=\pi$. Thus at short
distances, the running value of $\theta$ approaches $\pi$, and the
theory approaches the critical (conformal) field theory or RG
fixed-point theory of the transition itself.

The argument just given for the form of the confined pairs of
kinks off the transition is not rigorous as it stands, because it
is not clear that small regions of vacua other than the two on
either side of the transition (formed by allowing the kinks to
cross, or by creating a pair of kinks in the ``wrong'' ordering)
can really be neglected. It appears strongest in the second-order
case, where we may focus on low energies, and the spinons are
massless (gapless) at the critical point, but could fail
completely in the first-order case. To reinforce the argument for
the second-order case, we appeal to the lattice models, the six-
and supersymmetric- vertex models, or quantum spin chains. As
mentioned in subsection \ref{ssec:sigmod}, these are argued to be
in the same universality classes as the sigma models. In the spin
chain models (see eq.\ (\ref{tlham})), the two phases are at
$\epsilon=1+0^+$ and $1-0^+$ (our conventions imply that
$\theta=\pi - O(\ln \epsilon)$). In the ground state for
$\epsilon>1$, the expectation $\langle e_i \rangle$ for $i$ even
is larger than for $i$ odd, and {\it vice versa} for $\epsilon<1$.
This phenomenon is known as dimerization or spin-Peierls ordering,
especially when it occurs spontaneously, breaking the reflection
symmetry of the chain about a site present when $\epsilon=1$. Such
spontaneous symmetry breaking occurs in these models only for
$m>2$, implying that the transition as $\epsilon$ passes through 1
is first order. The phase when $\epsilon>1$ can be pictured as
represented by the state in which pairs of sites $i$, $i+1$ with
$i$ even form singlets. This is an eigenstate of $H$ only when
$\epsilon$ goes to $\infty$, but is a useful picture generally.
The other phase, the ground state for $\epsilon<1$, is represented
by dimerizing the other way, as singlet pairs $i$, $i+1$, $i$ odd.
Then a kink between these two phases must consist of a site that
is not forming a singlet pair with either neighbor, and so carries
the representation $V$ or $V^\ast$ exactly as for the sigma model
described above. This conclusion is exact, and not just an
artifact of the simple picture of the two phases. In the time
evolution governed by $H$, the action of the TL generators $e_i$
ensures that the unpaired sites never cross; their worldlines are
simply the lines in the graphical representation of the TL algebra
that we have already used. Thus the excitations in the critical
theory also consist of kinks that must alternate in type along the
system, and this is in fact implicit in the results in Ref.\
\cite{RS}. Off criticality, the kinks, which correspond to the
spin-1/2 kinks of the $U_q({\rm sl}_2)$ representation, are
confined in pairs to form the spin-1 kinks.

These arguments mean that in the continuum (scaling) limit, the
$S$-matrix of these spin-1 kinks in the sigma model off of
$\theta=\pi$ is of the form (\ref{Scz}). The matrices $E_i$, $X_i$
are written in terms of the TL generators as in (\ref{EXTL}), and
here the TL generators are in the supersymmetric representation
given by (\ref{TLsusyeven}), (\ref{TLsusyodd}). To repeat what was
stated in subsection \ref{ssec:sigmod} \cite{glr,RS}: in terms of
the alternating spaces $V$, $V^\ast$ for the kinks (with $V$ for
$i$ even, $V^\ast$ for $i$ odd), $e_i/m$ is the projector onto the
singlet in $V\otimes V^\ast$ (or $V^\ast\otimes V$) for neighbors
$i$, $i+1$. The expressions in eq.\ (\ref{EXTL}) are for the phase
$\theta=\pi+0^+$, or $\epsilon<1$, while those for the other phase
$\theta=\pi-0^+$, or $\epsilon>1$ are again obtained by adding
$-1$ to the indices on the $e_i$'s and ${\cal P}_i$'s. We note
again that for $m<\sqrt{3}$ (i.e. $m=1$!), the singlet part of a
bound kink pair in $V\otimes V^\ast$ (or $V^\ast\otimes V$) does
not correspond to a stable particle in the spectrum.


\section{The boundary Potts $S$-matrix, algebraically}

The scaling limit of the Potts model remains integrable in the
presence of certain boundary conditions on the half-plane.
In the language of Potts
spins, these are free and fixed boundary conditions, and we again
focus on the low-temperature phase. ``Free'' means the Potts spin
is unconstrained at the edge, and the model has the full $S_Q$
symmetry (though this is broken spontaneously in the low
temperature phase). ``Fixed'' means that the value of the Potts
spin at the edge is given, and the same all along the edge. In
this case, the symmetry is broken explicitly to $S_{Q-1}$. The
corresponding boundary $S$-matrices were found in \cite{Chim}. A
boundary $S$-matrix in an integrable theory must satisfy a variety
of constraints, including the boundary version of the YB equation
\cite{GZ}. The boundary YB equation is
$$R(u_2)S_i(u_1+u_2) R(u_1) S_i(u_1-u_2)
= S_i(u_1-u_2) R(u_1) S_i(u_1+u_2) R(u_2),$$ where $i=0$ or $N-2$
refers to the bulk $S$-matrix for the two bulk particles closest
to the boundary at the end. In the case at hand, the
boundary YB equation gives the overall form of $R$, the boundary
$S$-matrix. Since a solution of the boundary YB equation can be
multiplied by any function, one must use the remaining conditions
(the boundary analogs of the crossing, unitarity and bootstrap
conditions) to fix this function. In this section we show how to
formulate the two boundary $S$-matrices algebraically, so that the
boundary $S$-matrix applies to all the representations of the
Potts $S$-matrix.

The boundary $S$-matrix for fixed boundary conditions is proportional
to the identity matrix \cite{Chim}, and so is trivial to describe
algebraically. The reason for this is easy to see in the language of
Potts spins: since the spin on the boundary remains fixed, a kink
approaching the boundary must merely bounce off. Note that in this
case, with the Potts spin at one end of the system fixed, there are
exactly $(Q-1)^N$ states for $N$ kinks with given rapidities. This
space of states is a subspace (determined by the projectors ${\cal
P}_{2i+1}$, $i=0$, \ldots, $N-1$) of the states in which the $e_i$
($i=0$, \ldots, $2N$) act once more in the Potts representation
(\ref{pottsrep}), and the bulk $S$-matrices are given in terms of
$E_i$, $X_i$, by eq.\ (\ref{EXTL}), with $i=0$, \ldots, $N-2$ for $N$
kinks. We note that use of this representation of the $e_i$'s implies
that the total space contains $Q^{N+1}$ states, of which $Q(Q-1)^N$
survive projection, and the factor of $Q$ means that all possible
boundary spin values are in fact present, though they do not mix in
the scattering.

The boundary $S$-matrix for free boundary conditions is not
diagonal. In the language of Potts kinks, a kink $|ab\rangle$
scattering off the boundary at the right can scatter to any state
$|ab'\rangle$, as long as $b'\ne a$. The permutation symmetry
$S_Q$ means that the boundary $S$-matrix can be written as
\begin{equation}
R_{\rm free}(u)\propto I + r(u) Z, \label{Rfree}
\end{equation}
where in the Potts-kink representation used in \cite{Chim}, each element
$$Z_{a'b',\,ab}=\delta_{aa'},$$ i.e.\ in this representation, for each $a=a'$,
$Z$ is a $(Q-1)\times (Q-1)$ matrix with every entry $1$. Notice
that in the free case, the number of states of $N$ kinks with
given rapidities is $Q(Q-1)^N$, the same as in the fixed case, but
now all are mixed by the scattering processes.

This boundary $S$-matrix for the free case can be described in
terms of the TL generators. Precisely, if the $N$th kink is the
last before the boundary at the right, then we have
\begin{equation}
Z = m {\cal P}_{2N-1} e_{2N} {\cal P}_{2N-1}. \label{Zdef}
\end{equation}
The ${\cal P}_{2N-1}$ project the incoming particle and outgoing
particle onto $Q-1$
states, as required. The $e_{2N}$ then is responsible for the
boundary scattering. There is a similar expression for scattering
of the first kink off the boundary at the left.

\begin{figure}
\begin{center}
\leavevmode \epsfxsize 0.12\columnwidth \epsffile{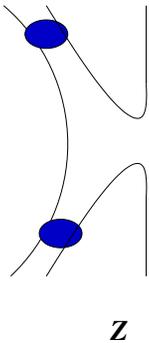}
\end{center}
\caption{Boundary scattering in terms of the TL generators}
\label{fig:bdry}
\end{figure}
In terms of the representation of the TL algebra in terms of
non-crossing lines, this can be represented as in figure
\ref{fig:bdry}.  There is a degree of freedom on the boundary,
represented by the vertical line, corresponding in the unrestricted
$U_q({\rm sl}_2)$ representation to the spin-1/2 representation. With
one of these at each end, they contribute the factor $m^2=Q$ to the
count of the number of states. The algebraic relations of $E_{N-2}$
and $X_{N-2}$ with $Z$ then follow immediately from the TL algebra
(\ref{tlrel}). There are many of them, so we will not write out them
all. The ones useful in solving the boundary YB equation are
\begin{eqnarray*}
Z^2 &=& (Q-1) Z,\cr E_{N-2} Z X_{N-2} &=& (Q-2) E_{N-2} Z,\cr Z
E_{N-2} Z E_{N-2} &=& (Q-1) Z E_{N-2},\cr Z E_{N-2} Z X_{N-2} &=&
(Q-2) Z E_{N-2} Z + Z E_{N-2},\cr Z X_{N-2} Z E_{N-2} &=&
(Q^2-3Q+1) Z E_{N-2}, \cr Z X_{N-2} Z X_{N-2} &=& (Q-2)^2 Z
E_{N-2} Z + (Q-2)ZE_{N-2} + ZX_{N-2}.
\end{eqnarray*}
Using these relations, it is then straightforward but tedious to
plug the $S$-matrix and $R$-matrix into the boundary YB equation
and find that
$$r(u) = \frac{\sinh(2\lambda u + 3i\gamma)}{2\cosh(\gamma)
\sinh(2\lambda u)}.$$ This is in agreement with the results of
\cite{Chim}, where the boundary $S$-matrix is written in the form
$$R_{\rm free}(u) = r_1(u) I + r_2(u) (Z-I),$$ so that $r=r_2/(r_1-r_2)$.
The boundary YB equation thus determines the ratio $r_1/r_2$, but
by using the boundary versions of crossing, unitarity, and the
bootstrap, the functions themselves have been determined \cite{Chim}.

The utility of the expression (\ref{Zdef}) is that the boundary
$S$-matrix can now be found explicitly in any representation of
interest. In addition to the Potts-kink representation of
\cite{Chim}, there is the high-temperature version, in which $-1$
is added to the indices in the expressions for $E_i$, $X_i$ and
$Z$. The $S$-matrix for the ``fixed'' boundary condition of the
low-temperature case becomes that for the free boundary condition
of the high-temperature case, and can be represented in a total
space of only $Q^N$ states (before application of the projectors).
The $S$-matrix for the ``free'' boundary condition of the
low-temperature case becomes one with additional boundary degrees
of freedom (discussed further below) in the high-temperature case,
and presumably means a fixed boundary condition on the {\em dual}
Potts variables. The $U_q(A^{(2)}_2)$ representation of the
boundary $S$-matrix was found in \cite{kim}. The RSOS-kink
representation (valid for all the $\Phi_{21}$-perturbed minimal
models) and supersymmetric representation of this boundary
$S$-matrix seem to be previously unknown. One can easily extend
these considerations to the $\Phi_{12}$-perturbed minimal models
as well.

Once again, the application in the example of the ${\bf
CP}^{m+n-1|n}$ sigma models deserves comment. In the presence of a
boundary, the topological term in the sigma model is no longer an
integer, and the physics is not periodic in $\theta$ (this and the
effect of the $\theta$ term on the boundary conditions are
discussed extensively in Ref. \cite{xrs}). The region
$-\pi<\theta<\pi$ behaves like the $\epsilon>1$ region of the
quantum spin chain with Hamiltonian (\ref{tlham}). In this case
the dimerization extends up to the end of the chain, and there is
no boundary degree of freedom. On the other hand, in the region
$\pi<\theta<3\pi$, the spin chain dimerizes the other way, and an
unpaired spinon is left at each end; the spinon (kink) in the
representation $V$ is bound by a linear potential to the left end
of the system, and another in the dual $V^\ast$ is bound at the
right end (if the system is large). These occur to lower the bulk
energy density of the vacuum to its lowest possible value
\cite{oldCPN}; the boundary ``knows'' that the underlying $\theta$
is $>\pi$ and so a different vacuum occurs immediately adjacent to
the edge, separated from the bulk by a kink. Consequently, with
$N$ particles (bound kink pairs) in the bulk, the number (i.e.
dimension of the space) of states for a given set of rapidities is
$(m+2n)^2[(m+2n)^2-1]^N$, similar to the Potts model with $Q=m^2$
[however, the superdimension of the space \cite{RS} is
$m^2(m^2-1)^N$]. Similarly, if we go to the region $|\theta-2\pi
s| <\pi$, $s\geq0$, then there will be $s$ spinons bound at each
end. For $s\leq0$, the role of $V$ and $V^\ast$ at the boundary is
reversed, as well as in the bulk.

Explicitly, for the scaling limit of the spin chain with $V$,
$V^\ast$ alternating, such that $V$ is at the left end, and
$V^\ast$ is at the right, we find that for $\epsilon>1$,
corresponding to $\theta<\pi$, the particles in the bulk have $V$
to the left of $V^\ast$ (see end of section 3), and there is no
degree of freedom on the boundary, so the forms for the
high-temperature Potts phase with free boundary conditions apply.
For the spin chain with the same boundary conditions as before but
now with $\epsilon<1$ ($\theta>\pi$), the particles in the bulk
have $V^\ast$ to the left of $V$, and there is a boundary spinon
in $V$ at the left, and $V^\ast$ at the right. In this case the
$S$-matrices are precisely as given in eqs.\ (\ref{EXTL}) and
(\ref{Zdef}), the free boundary conditions for the low-temperature
Potts phase.

We have found that the cases $s=\pm1$, as well as $s=0$, are
integrable in the scaling limit near the transition, for $m=2$. We
do not know if the cases $|s|>1$ are integrable (the spin chain
models with larger $\cal S$ contain phases with $s>1$ as well as
$s=0$, $1$, or negative values if one spin is added or removed
from each end of the chain).

It is amusing that in the application of the ${\bf CP}^{1|1}$
sigma model to the spin quantum Hall transition, the boundary
degree of freedom present on one side of the transition is just
the famous ``edge state'' familiar from the related integer
quantum Hall effect. Thus here we have, in principle, access to
some exact properties of these states in the scaling regime near
the transition.


\section{Other $S$-matrices for supersymmetric sigma models}
\label{sec:disc}

In the previous sections, we have obtained the $S$-matrices for a
perturbation of the critical theory of the ${\bf CP}^{m+n-1|n}$
sigma models at $\theta=\pi$ by a change in $\theta$, which
corresponds to an RG flow from the critical theory at $\theta=\pi$
in the ultraviolet to the $\theta=0$ massive fixed point in the
infrared. A byproduct is that if we take the ultraviolet limit of
our $S$-matrices, or equivalently let the mass scale $M$ go to
zero with the energies and momenta of the particles fixed, then we
obtain a description of the critical theory at the $\theta=\pi$
fixed point. In fact, we obtain two inequivalent, but dual
descriptions of this theory, since our particles either have the
representation $V$ to the left of $V^\ast$, or vice versa,
depending which side of the transition we come from. These
representations $V$, $V^\ast$ are dual and inequivalent (except in
the cases $n=0$ and either $m=1$, which is trivial, or $m=2$,
which is the usual SU$(2)_1$ WZW theory), and hence so are our
descriptions of the theory. Presumably though, they do lead to
descriptions of the same theory in the ultraviolet.

Other $S$-matrices are known for certain sigma models. In particular,
for the O$(3)$ sigma model, there is an $S$-matrix for massive
particles in the vector (spin 1) representation of SO$(3)$ [or even
for SO$(m)$], which represents the RG flow from the weak-coupling
fixed point in the ultraviolet to the strong-coupling fixed point at
$\theta=0$ (mod $2\pi$) in the infrared \cite{ZZ1}. There is also
another for {\em massless} spin-1/2 particles that represents the RG
flow from weak coupling into the nontrivial fixed point at
$\theta=\pi$ in the infrared \cite{ZZ2}. This one is for the O$(3)$
model only (the O$(m)$ model for $m>3$ has no $\theta$ term). Both of
these possess generalizations in which the symmetry is deformed to
$U_q({\rm sl}_2)$. In the RSOS representations, these have been used
to describe parafermion theories under an integrable perturbation
\cite{Fateev}.

For symmetry reasons, both of these $S$-matrices can be written in
terms of the TL algebra. Thus following the work
of section 4, we can use the supersymmetric representation
of the TL algebra to find the supersymmetric $S$-matrices.
This is simple to see for the massless spin-1/2
($\theta=\pi$) case. Here the $S$-matrix is a linear combination
of $I$ and $e$, with $u$-dependent coefficients. There are
different $S$-matrices for scattering left- with left-, right-
with right-, or left- (right-) with right- (left-) -moving
particles. We may now utilize these $S$-matrices in the
supersymmetric representation of the TL algebra. The massless
particles in the spectrum at $\theta=\pi$ are assumed to be kinks
that are alternately in the $V$ or $V^\ast$ representations. Then
it is immediate that the $S$-matrices for these massless kinks can
be written in terms of the TL
generators.

For the spin-1 representation of $U_q({\rm sl}_2)$, the $S$-matrix
can be written in terms of the generators $I$, $E$, $X$ of the
SO$(3)$ BMW algebra. As in previous sections, we can then use the
same forms (with the BMW generators written in terms of TL
generators as before) for scattering of bound pairs of kinks,
where the kinks must alternate, and so either every pair has $V$
to the left of $V^\ast$, or every pair has the reverse order.

These constructions give $S$-matrices in supersymmetric
representations that parallel those for the O$(3)$ sigma model RG
flows at $\theta=0$ or $\pi$, respectively. It is natural to try
to identify these $S$-matrices as describing the analogous flows
in the ${\bf CP}^{m+n-1|n}$ sigma models, for $0<m\leq 2$ where
similar RG flows are expected. Certainly the types of particles
present in the infrared limit seem plausible: kinks in $V$,
$V^\ast$ alternately in the $\theta=\pi$ case, and in the adjoint
(part of $V\otimes V^\ast$) for $\theta=0$. Actually, the latter
is disturbing, since as we pointed out, there are two distinct
forms, depending whether $V$ is to the left or right of $V^\ast$
in all the particles, and this appears to give two distinct
$S$-matrices. The choice of one or other of these would break the
reflection (parity) symmetry of the model at $\theta=0$. It could
be that there is an isomorphism that would show the equivalence of
the two choices, for particular values of $m$, but this is not
clear to us. (It was not an issue at all in the previous sections,
where reflection symmetry was explicitly broken by $\theta\neq 0$
or $\pi$, nor for the $\theta=\pi$ case here with free boundaries,
since $\theta=\pi$ and $-\pi$ are then distinct.) A check on the
proposal is to calculate properties in the ultraviolet. From the
thermodynamic Bethe ansatz, the central charge of the theory at
the fixed point in the ultraviolet can be obtained. It is known
that this is $c=2(m-1)$ for $m=1$, $2$, for either case $\theta=0$
or $\pi$ (the calculations are thermodynamic and independent of
the representation used, and hence independent of $n$). The
correct value at the weak-coupling fixed points in the ${\bf
CP}^{m+n-1|n}$ sigma models is always $c=2(m-1)$, which is a
useful check on our claims.

It is possible to construct spin-chain models, different from
those in Sec.\ \ref{ssec:sigmod}, that are described in the
continuum limit by these supersymmetric $S$-matrices. The spins
have spin ${\cal S}$ under the $U_q({\rm sl}_2)$ symmetry.  To
construct a Heisenberg-like coupling between nearest neighbors, we
need to study their $U_q({\rm sl}_2)$ representation properties.
In general, two spins decompose as in ${\rm sl}_2$, into total
spins $0$, $1$, \ldots, $2{\cal S}$ .  This can always be done
within the TL algebra, by projecting of a set of $2{\cal S}$
spin-1/2's onto their spin-${\cal S}$ part. (In the ${\cal S}=1$
case, one of course ends up with the BMW algebra, as has been
discussed at length in this paper.)  Then we may take a linear
combination of these terms chosen so that the possible eigenvalues
in each total spin sector are those of the invariant bilinear form
of generators of $U_q({\rm sl}_2)$, with one generator on each
site. This is then the analog of Heisenberg coupling for $U_q({\rm
sl}_2)$, and can be carried over to any representation of the TL
algebra, such as the supersymmetric ones.  With this we may then
build a chain out of these ``spin-${\cal S}$'' representations in
any representation of the TL algebra, by using this coupling for
nearest neighbors. It consists of $2N$ medial-graph sites, with
$2N$ divisible by $2{\cal S}$. Note that for $2{\cal S}$ odd, the
spin-${\cal S}$ representations must alternate with their duals,
in order that the TL sites always alternate between $V$ and
$V^\ast$ in the supersymmetric representation. For $2{\cal S}$
even, there are two different constructions, depending whether in
each group of sites $V$ or $V^\ast$ is the first one at the left.
We again consider only free boundary conditions. Notice that the
spin-${\cal S}$ representations in the supersymmetric
representation of TL are multiplets with the degeneracies found in
the appendix of Ref.\ \cite{RS}, which are larger than those of
irreducible representations of sl$(m+n|n)$ for $n>0$, except for
the cases ${\cal S}=0$, $1$, which are the singlet and adjoint of
sl$(m+n|n)$. Notice also that the ``spin-${\cal S}$''
representations here, which are part of the product $V\otimes
V^\ast\otimes V \cdots$ ($2{\cal S}$ times), [or $V^\ast\otimes
V\otimes V^\ast \cdots$ ($2{\cal S}$ times), or these alternately
if $2{\cal S}$ is odd], are different from the ``spin-$\cal S$''
representations in Sec.\ \ref{ssec:sigmod}, which alternated
between supersymmetrized products of $2{\cal S}$ copies of $V$ and
of the same for $V^\ast$.

Like the spin-${\cal S}$ $U_q({\rm sl}_2)$ Heisenberg chain they
are related to, these lattice models are not integrable, except
for ${\cal S}=1/2$. (It is possible to fine-tune the coefficients
of the nearest-neighbor couplings to obtain integrable models, but
these are multicritical and not our interest here. Such models
have been constructed by this same procedure in Ref.\ \cite{ks},
though the supersymmetric representation was not considered.) But
as ${\cal S}\to\infty$, a semiclassical approximation is valid, as
for the SU$(2)$ ($q=1$) chains \cite{Haldane}. In such a limit,
this chain will map onto some sort of
nonlinear sigma model. For the supersymmetric representation of TL
with $n>0$, the target space of the sigma model is a
noncommutative space, that so far has been defined only by the
``space of functions'' on it, which by definition consists of the
multiplets of total ${\cal S}=0$, $1$, $2$, \ldots, which occur in
the decomposition of the product of two sites, for $2S$ either
even or odd, as ${\cal S}\to\infty$. By construction, the
eigenvalues of the Hamiltonians of these chains can be found for
finite length and finite ${\cal S}$ using only the TL algebra, and
so are independent of the representation used (though
multiplicities could vanish in particular cases). In the continuum
${\cal S}\to\infty$ limit (taken in the obvious way, with the
lattice constant going to zero, and the velocity of ``light'', the
mass scale $M$, and the length $L$ of the system staying fixed) we
then expect that these theories {\em are} integrable, and we have
no doubt that the $S$-matrices for all the RG flows considered in
this paper, including this section, apply to these models. In
particular, the lines in the graphical representation of TL never
cross, as in the $S$-matrix constructions, and this is related to
the symmetry properties in Ref.\ \cite{rsunpub}. It is not obvious
that there is a problem with parity symmetry, as there are two
distinct models microscopically for $\theta=0$ (we will use the
same terminology as before, $\theta=0$ and $\theta=\pi$ for the
cases with no staggering of the couplings and ${\cal S}\to\infty$
through values $2{\cal S}$ even and odd, respectively).

A more difficult question is whether these $S$-matrices describe
the ${\bf CP}^{m+n-1|n}$ sigma models in their flows along
$\theta=0$ or $\pi$, or in other words, whether the continuum
limits of our spin chains are the same as these sigma models. In
view of the different geometry of the target spaces in the
semiclassical (weak-coupling) limit, this seems unlikely. In
particular, we may compare the spectra of the continuum models in
the weak-coupling/semiclassical regime in finite size, where the
length acts as an infrared cutoff. For free boundary conditions,
the lowest states are those where the sigma model field is
constant along the system, and behaves as a free
quantum-mechanical particle moving on the target space. For both
the models defined here, and the ${\bf CP}^{m+n-1|n}$ sigma
models, these lowest states can be labelled in increasing order by
total ${\cal S}=0$, $1$, $2$, etc, where ${\cal S}=0$, $1$ are the
singlet and adjoint of sl$(m+n|n)$, but for higher ${\cal S}$ the
multiplicities in the models based on the TL algebra are those
found in the appendix of Ref.\ \cite{RS}, and are larger than
those for motion on ${\bf CP}^{m+n-1|n}$. The latter can be found
from the spin chain models in Sec.\ \ref{ssec:sigmod}, by
decomposing two sites of spin-${\cal S}$ (as defined there), and
taking ${\cal S}\to\infty$. In fact, for given $m$, $n$, the
multiplicities increase exponentially with $\cal S$ in the models
here \cite{RS}, but as powers of $\cal S$ for the ${\bf
CP}^{m+n-1|n}$ sigma models, or the spin chain models of Sec.\
\ref{ssec:sigmod}. On the other hand, for a long system $L\gg
M^{-1}$, the states at finite excitation energy are built from a
finite number of excitations (massive particles) in the adjoint
(for the $\theta=0$ case) in both sigma models, and it is not
obviously ruled out that the spectra coincide.

We should point out that the partition functions, with no
supersymmetry-breaking source terms, and with supersymmetric
boundary conditions, for the ${\bf CP}^{m+n-1|n}$ sigma models and
for the continuum models constructed here (with weak coupling in
the ultraviolet) are in both cases the same as for $U_q({\rm
sl}_2)$ continuum models, which become the O$(3)$ sigma model for
the $q=1$ ($m=2$) case. This is a consequence of the
supersymmetry, which leads to a cancellation of corresponding
bosonic and fermionic states in the familiar way. In general, it
does not imply that all energy levels of theories that have equal
partition functions are the same (if ``partition function'' means
${\rm STr\,}e^{-\beta H}$ \cite{RS} for supersymmetric
formulations, and the $q$-deformed trace for the $U_q({\rm sl}_2)$
formulation), since some may occur in multiplets of vanishing
super- (or $q$-) dimension that do not contribute to the
supertrace STr (resp., $q$-deformed trace). However, for the spin
chains constructed here within the TL algebra, all the energy
levels for different models do coincide, at least in the $m=2$
case, since the spin-1/2 (XXZ or TL) representation of the TL
algebra is faithful, as is the supersymmetric one (for $n>0$), and
this result remains true in the continuum limit. For the spin
chains of Sec.\ \ref{ssec:sigmod}, which lead to the ${\bf
CP}^{m+n-1|n}$ models in the continuum limit, it is not clear to
us if all the levels are in fact the same as those in the other
two models for each $m$.

While the continuum models constructed here are integrable, the
standard arguments for integrability break down in the ${\bf
CP}^{m+n-1|n}$ sigma model with $n>0$ for $m\neq 2$ \cite{salpr}.
Thus it seems unlikely that the $m=1$ case of these flows or spin
chains agrees with the ${\bf CP}^{m+n-1|n}$ sigma models. In other
words they are different theories, even though the partition
functions coincide. For $m=2$, one suspects that the general
$n>0$ model is integrable, because of the usual arguments that the
physics is the same for all $n$, and that its spectrum,
$S$-matrices, etc, are related to those of the $q=1$ SU$(2)$
model. Thus some kind of limited equivalence of these
sl$(m+n|n)$-invariant theories is not ruled out. This leaves us
uncertain how to describe the relation of the different theories,
and whether all are integrable to the same extent. For the fixed
point at $\theta=\pi$, and as a consequence also for the flow out
of it, we believe that the theories coincide, as the spin chain
models for ${\cal S}=1/2$ do, so the results of earlier sections
are not affected.

We may note in passing that there are also natural constructions
of boundary $S$-matrices with a spin-${\cal S}$ on the boundary,
for any ${\cal S}$, in the models introduced in this section.
The boundary $S$-matrices would be found by fusion, as described
for the Kondo problem in \cite{pkondo}.
These would naturally describe different boundary conditions on
the sigma model, or the spin chains depending on the ${\cal S}$ at
each site of the chain and the degree of staggering of the nearest
neighbor couplings. However, except when the boundary ${\cal S}$
is 0 or $1/2$, these representations of sl$(m+n|n)$ are not those
expected in the ${\bf CP}^{m+n-1|n}$ sigma models, as discussed in
the previous section. The representations in the two cases are the
same as those for the respective spin chains.

These $S$-matrices now exhaust the $U_q({\rm sl}_2)$-invariant
$S$-matrices known to us that could be used in different
representations of the TL algebra. On the other hand, there are
representations of the SO$(3)$ BMW algebra that do not arise from
representations of the TL algebra. A case in point is where the
strands in figure \ref{fig:bmw} depict the vector representation
of OSp$(3+2n|2n)$, or more generally of $U_q({\rm osp}(3+2n|2n))$.
This is the basic construction of the SO$(3)$ BMW algebra,
extended to orthosymplectic supersymmetry \cite{sw}. For example,
for $q=1$, it is related to SU$(2)_1$/4-state--Potts. Similarly,
we could use representations of the TL algebra with $U_q({\rm
sl}(2+n|n))$ supersymmetry. In both cases, the properties will be
related to the $n=0$ cases as in earlier examples in this paper.


\section{Conclusion}

We have found $S$-matrices for the supersymmetric sigma models
with $m=1$, $2$ that correspond to the flow from the fixed point
at $\theta=\pi$ to the massive theory at $\theta=0$. These results
have potential applications to the spin quantum Hall transition in
disordered fermion systems.

One can use the $S$-matrix to compute various physical quantities.
For example, universal amplitude rations for the Potts models were
computed in \cite{Delfino}. The thermodynamic Bethe ansatz
computation for the Potts field theory was done in \cite{Dorey2},
but unfortunately the results are valid only at integer values
$p\ge 3$, and so cannot be applied to the percolation limit $Q\to
1$. Perhaps our results will be useful in extending these
computations further.

One interesting open question is to understand the relation between
the two types of sigma models discussed in section 6. These models seem
to be thermodynamically equivalent: does that mean that they have the
same $S$-matrix?

\vspace{0.35in}

We thank H. Saleur for useful comments. NR is grateful to Andreas
Ludwig for early discussions of the massless $S$-matrices at
$\theta=\pi$ in section \ref{sec:disc}.
This work was supported in part by NSF
grant DMR-0104799, a DOE OJI award, and a Sloan Foundation Fellowship
(P.F.), and NSF grant DMR-98-18259 (NR).


\appendix
\section{SU$(m)$-invariant $S$-matrices for $m>2$?}

In this appendix we consider a question that arises naturally from
the discussion in section 4: can we find exact $S$-matrices for
the massive kinks in the $m>2$ cases, assuming they always
alternate kink and anti-kink? These might resemble exact
$S$-matrices for the ${\bf CP}^{m-1}$ sigma model (if we set
$n=0$), however, similar questions to those raised in Sec.\
\ref{sec:disc} apply here as well. The ${\bf CP}^{m-1}$ sigma
model is usually thought (though not proven) not to be integrable
at $\theta=0$ if $m>2$; for $\theta=\pi$, we know of no conclusive
arguments either way.  By continuing the earlier results to $m>2$,
we find seemingly-sensible $S$-matrices, but they have a peculiar
periodicity in the rapidity that rules them out from describing
the sigma models. (Conceivably, at $\theta=\pi$, the assumption
that the kinks and anti-kinks alternate is incorrect when the
transition at $\theta=\pi$ is first-order.) Nonetheless, some of
these $S$-matrices contradict a ``proof'' of \cite{BW} that there
is no $S$-matrix for particles in the adjoint of SU$(m)$ for
general $m$.

Since $m>2$, we can set $n=0$ in this appendix (the results generalize
easily to $n>0$).  We first study the case where particles are in the
fundamental representation of SU($m$), and the particles alternate in
space between the fundamental representation $V$ (denoted here as
$m$), and the dual $V^\ast$ (denoted $\overline m$). The two-particle
$S$-matrix in this model therefore is between a particle in the $m$
representation and one in the $\overline{m}$, which we denote
$S_{m\overline{m}}=S_{\overline{m}m}$, as it is independent of which
is which. Because the particles must always alternate between $m$ and
$\overline{m}$, there is no $S_{mm}$ or $S_{\overline{m}\overline{m}}$
in this model.
In order to preserve the SU$(m)$ symmetry, we must have
$$S_{m\overline{m}}(u) \propto f^{(0)}(u){\cal P}^{(0)}+
f^{(a)}(u){\cal P}^{(a)},$$ where ${\cal P}^{(0)}$ and ${\cal
P}^{(a)}$ project the tensor product of the $m$ and $\overline{m}$
onto the singlet and adjoint representations. In section 3 we saw
that these operators are $${\cal P}^{(0)}=\frac{e}{m},
\qquad\qquad {\cal P}^{(a)}=1-\frac{e}{m}.$$ Using this, we can
rewrite the $S$-matrix as
\begin{equation}
S_{m\overline{m}}(u) = Z(u) \left(I + F(u) e \right). \label{SNN}
\end{equation}

The function $F(u)$ is found by demanding $S_{m\overline{m}}$ that
satisfy the YB equation, which here is
$$\left(I+F(u)e_i\right)\left(I+F(u+u')e_{i+1}\right)
\left(I+F(u')e_i\right)=
\left(I+F(u')e_{i+1}\right)\left(I+F(u+u')e_i\right)\left(I+F(u')
e_{i+1}\right).
$$
Using the  TL relations (\ref{tlrel}), one finds that
$$F(u)+F(u')+NF(u)F(u') + F(u)F(u+u')F(u') - F(u+u') = 0$$
The only non-trivial solution of this functional equation is
\begin{equation}
F(u) = -\frac{\sinh[Au]}{\sinh[\pi\widetilde{\gamma}+Au]},
\label{forF}
\end{equation}
where
$$N=2\cosh(\pi\widetilde{\gamma}),$$
and the parameter $A$ is not constrained by the YB equation. Note
that we have defined $\widetilde{\gamma}$ so that it is related to
our earlier parameter $\gamma$ by
\begin{equation}
\pi\widetilde{\gamma}=i\left(\gamma-\frac{\pi}{2}\right).
\label{gamgam}
\end{equation}
Before, we were interested in $m\le 2$, where $\gamma$ was real.
Now we are interested in $m > 2$, where $\widetilde{\gamma}$ is
real. This is a familiar solution of the YB equation when $\gamma$
is real: it is that associated with  the sine-Gordon $S$-matrix
and the O$(m)$ model $S$-matrix for $m<2$ (as explained carefully
in \cite{Zpoly,Smirpoly}). In lattice-model language, when
$\gamma$ is real this is the solution of the YB equation
associated with the six-vertex model and the RSOS models. Thus
what we have done here is argue that $S_{m\overline{m}}$ is
essentially the continuation of a very familiar $S$-matrix to
$m>2$.

To complete the determination of the $S$-matrix, we need to find
the function $Z(\theta)$ and the constant $A$.  Unitarity requires
that $S_{m\overline{m}}(u) S_{m\overline{m}}(-u)=I$, so
$$Z(u)Z(-u)=1.$$
Crossing symmetry requires that $S$-matrix elements are related to
those ``rotated'' by $90$ degrees, namely
$$S(a\overline{b} \to c\overline{d})(u) =
S(\overline{b}d \to \overline{a}c)(i\pi-u).$$
The simplest way of satisfying this relation is to have
\begin{eqnarray*}
A&=&-i\pi\widetilde{\gamma} + j,\cr Z(u)&=&Z(i\pi -u)
\frac{\sin\left[\widetilde{\gamma}(i\pi-u)\right]}
{\sin[\widetilde{\gamma}\theta]},
\end{eqnarray*}
where $j$ is some integer. We set $j=0$, because a non-zero $j$
will mean that the $S$-matrix is no longer unitary.
The simplest $Z(\theta)$ obeying these relations is then
\begin{equation}
Z(u)=\prod_{p=1}^{\infty}
\frac{\sin\left[\widetilde{\gamma}((2p-1)i\pi -u)\right]
\sin\left[\widetilde{\gamma}(2pi\pi +u)\right]}
{\sin\left[\widetilde{\gamma}((2p-1)i\pi +u)\right]
\sin\left[\widetilde{\gamma}(2pi\pi - u)\right]}.
\end{equation}
Note that $A$ is purely imaginary, so that the $S$-matrix has
periodicity under $u\to u+2\pi/\widetilde{\gamma}$. Most $S$-matrices
are periodic under {\it imaginary} shifts of rapidity, not real ones.
The function $Z(u)$ is an elliptic function: in addition to the
periodicity $Z(u)=Z(u+2\pi/\widetilde{\gamma})$, it also has a
quasi-periodicity under imaginary shifts in $u$, namely
$$Z(u+2i\pi) = Z(u)\frac{\sin^2[\widetilde{\gamma}(i\pi +u)]}
{\sin[\widetilde{\gamma}u]\sin[\widetilde{\gamma}(u+2i\pi)]}.$$
One can presumably use this to rewrite $Z(u)$ in terms of the
usual elliptic $\theta$ functions.

An $S$-matrix periodic in real rapidity is unusual, to say the
least. Such $S$-matrices have been discussed several times in the
literature. In \cite{Zellip}, an $S$-matrix based on the $R$-matrix
for the lattice eight-vertex model was discussed. In \cite{Mussardo},
an elliptic generalization of the sinh-Gordon $S$-matrix was
discussed, and resulting form factors were computed.  In both of these
earlier cases, it was not clear which (if any) quantum field theories
possessed such $S$-matrices. Here, we were motivated by the ${\bf
CP}^{m-1}$ sigma model at $\theta=\pi$ to study a theory where
particles alternate between the $m$ and $\overline{m}$ representations
of $SU(m)$.  However, this $S$-matrix does not seem to describe the
sigma model.  If the massive kinks described here are the excitations
at $\theta=\pi$, then the ultraviolet limit of the $S$-matrix would
describe the weakly-coupled sigma model at $\theta=\pi$. This is
evidently not the case. Precisely, one can compute the energy in the
presence of a background field from the $S$-matrix, and from the
weak-coupling expansion sigma model of the sigma model. These two do
not agree; the weak-coupling perturbation expansion shows no evidence
of periodicity in the $S$-matrix.

It is expected that the ${\bf CP}^{m-1}$ model at $\theta=0$
exhibits confinement. This has been established for $m=2$ and
large $m$, and is believed to hold for all $m>0$ \cite{oldCPN}.
Confinement means that even though the fields are in the
fundamental representations, no particles in the fundamental
representations appear in the spectrum. Instead, the particles are
believed to be in the adjoint representation of SU$(m)$. This has
been proven for $m=2$, where it has long been known that the
particles are in the three-dimensional representation of SU$(2)$
\cite{ZZ1}.

One can use our results to construct an $S$-matrix for particles
in the adjoint of SU$(m)$. There is a claim in the literature that
the only such $S$-matrix must be O$(m^2-1)$ invariant; we will
show here that this is not so. As noted in section \ref{ssec:ssS},
with the appropriate representation of the TL algebra, the Potts
$S$-matrix describes scattering of particles in the adjoint of
SU$(m)$ for $m\le 2$. We have just shown that the O$(m)$ model
$S$-matrix for $m<2$ can be ``continued'' to $m>2$. The same goes
for the Potts $S$-matrix described in Sec.\ \ref{ssec:ssS}, or for
the spin-$1$ $U_q(\rm{sl}_2)$ $S$-matrix described in Sec.\ \ref{sec:disc}.
For the former, we consider an $S$-matrix of the same form as (\ref{Scz}),
namely
$$S_{\rm adj}(u)=Z_{\rm adj}(u)(g(u)I + h(u) E + X),$$
where $E$ and $X$ obey the algebraic relations (\ref{EXalg}),
(\ref{EXEalg}), with as always $Q=m^2$. Then, as discussed above,
$S_{\rm adj}$ obeys the YB equation if the functions $g(u)$ and
$h(u)$ satisfy (\ref{gh}), where $m=2\sin\gamma$.
However, this $S$-matrix is very different
from the Potts $S$-matrix because when $m>2$, $\gamma$ is no
longer real; it must be of the form (\ref{gamgam}) with
$\widetilde{\gamma}$ real. Now if we take $E$ and $X$ to be
defined in terms of the TL generator $e$ as in (\ref{EXTL}), and
we take the representation of the TL algebra as in the SU$(m)$
quantum spin chains, then $S_{\rm adj}$ indeed describes the
scattering of particles in the adjoint representation of SU$(m)$.
Such an $S$-matrix is not O$(m^2-1)$ invariant.

To complete this $S$-matrix, we need to find the function $Z_{\rm
adj}(u)$. By using unitarity and crossing, we find that
\begin{eqnarray*}
g(u)&=&\frac{\sin\left[\widetilde\gamma(2i\pi -3u)\right]}
{\sin\left[3\widetilde\gamma u\right]},\\
Z_{\rm adj}(u)&=&Z_{\rm adj}(i\pi - u),\\
Z_{\rm adj}(u)Z_{\rm adj}(-u)g(u)g(-u)  &=& 1.
\end{eqnarray*}
The simplest solution for $Z_{\rm adj}$ is therefore
\begin{eqnarray*}
&&Z_{\rm adj}(u)= \frac{1}{g(-u)}\times\\
&&\qquad\prod_{p=1}^{\infty}
\frac{\sin\left[\widetilde{\gamma}((6p-3)i\pi -u)\right]
\sin\left[\widetilde{\gamma}(6pi\pi +u)\right]
\sin\left[\widetilde{\gamma}((6p+2)i\pi -u)\right]
\sin\left[\widetilde{\gamma}((6p-1)i\pi +u)\right] }
{\sin\left[\widetilde{\gamma}((6p-3)i\pi +u)\right]
\sin\left[\widetilde{\gamma}(6pi\pi -u)\right]
\sin\left[\widetilde{\gamma}((6p+2)i\pi +u)\right]
\sin\left[\widetilde{\gamma}((6p-1)i\pi -u)\right] }.
\end{eqnarray*}
Like the $S$-matrix for particles alternating between $m$ and
$\overline{m}$ representations, this is periodic under real shifts
in rapidity and quasi-periodic under imaginary shifts.

The other SU$(m)$ $S$-matrices for particles in the adjoint, that
are the continuation of those in Sec.\ \ref{sec:disc} to $m>2$,
also exhibit the same peculiar periodicity. Our conclusion is
therefore that these $S$-matrices do not describe either the
$\theta=\pi$ or $\theta=0$ ${\bf CP}^{m-1}$ models. They join the
$S$-matrices of \cite{Zellip,Mussardo} as interesting $S$-matrices
with no known physical interpretation.


%

\end{document}